\documentclass[a4paper,11pt]{article}
\usepackage{jcappub} 
\usepackage{lineno}
\usepackage{comment}
\usepackage{multirow}

\usepackage{xcolor}
\usepackage{booktabs}
\usepackage{graphicx}
\usepackage{soul}
\usepackage{LB}

\providecommand{\simone}[1]{{#1}} 

\title{\boldmath Systematic effect induced by misalignment in a Reflective Polarization Modulator for CMB, and application to the LiteBIRD case}

\author[1,2,3]{S.~Stellati,}
\author[1,2]{F.~Piacentini,}
\author[1,2]{S.~Micheli,}
\author[1,4]{A.~Novelli,}
\author[1,2]{F.~Columbro,}
\author[1,2]{A.~Coppolecchia,}
\author[1,2]{P.~de~Bernardis,}
\author[1,2]{G.~Pisano,}
\author[1,2]{S.~Masi,}
\author[1,2]{M.~Najafi,}
\author[1,2]{A.~Occhiuzzi,}
\author[4,5]{L.~Pagano,}
\author[1,2]{A.~Paiella,}
\author[]{\\ for the LiteBIRD Collaboration}

\affiliation[1]{Dipartimento di Fisica, Univ. La Sapienza, P. le A. Moro 2, Roma, Italy}
\affiliation[2]{INFN Sezione di Roma, P.le A. Moro 2, 00185 Roma, Italy}
\affiliation[3]{Dipartimento di Fisica, Univ. di Tor Vergata, Via della Ricerca Scientifica 1, Roma, Italy}
\affiliation[4]{University of Milano Bicocca, Physics Department, p.zza della Scienza, 3, 20126 Milan, Italy}
\affiliation[5]{Dipartimento di Fisica e Scienze della Terra, Università di Ferrara, Polo Scientifico e Tecnologico – Edificio C Via Saragat, 1, 44122 Ferrara, Italy}
\affiliation[6]{Istituto Nazionale di Fisica Nucleare, sezione di Ferrara, Polo Scientifico e Tecnologico – Edificio C Via Saragat, 1, 44122 Ferrara, Italy}

\emailAdd{simone.stellati@uniroma1.it}

\abstract{The \LiteBIRD mission aims to measure the Cosmic Microwave Background (CMB) polarization with unprecedented precision to detect primordial $B$ modes and constrain the tensor-to-scalar ratio $r$. A key component of \LiteBIRD are the polarization modulators, based on Half-Wave Plates (HWP). In this work, we study the systematic effects caused by a constant misalignment between the HWP’s rotation axis and optical axis, in case of a reflective HWP, when the misalignment is small and thus not corrected for in the pointing reconstruction. While the HWP shape is assumed to be ideal (i.e., plane-parallel), this misalignment mimics the effect of a physical wedge. Throughout the paper, we refer to this as a wedge-like effect.
This analysis also applies to the case of transmissive HWP with non-parallel surfaces, and is quantified by a \textit{wedge} angle. 
This misalignment leads to HWP-synchronous pointing error. Using the \LiteBIRD simulation framework, we implement the wedge-like effect in time-ordered data (TOD) and analyze its impact on reconstructed maps and angular power spectra. We find that the wedge systematically bias polarization measurements, leading to spurious $B$ modes. 
By evaluating the total error \simone{induced by the wedge-like effect $\Delta r_{\text{wedge}}$} on the tensor-to-scalar ratio, we establish a constraint on the maximum allowable wedge angle ($\alpha_{\text{max}}$) to ensure systematic errors remain below mission requirements. Our analysis, based on CMB-only simulations with orthogonal detector pairs (ranging from two to six detectors in total), assuming perfectly matching detectors within each pair and neglecting noise and foregrounds, provides a constraint of $\alpha_{\text{max}} = 22.7$\,arcmin from the point of view of spurious polarization. Beam reconstruction and pointing systematics are anyway dominant for values of $\alpha$ larger than 4\,arcmin. Interestingly we find that the contamination mimics lensing $B$ modes, and not primordial tensor modes and that it is reduced increasing the number of detectors. 
This study highlights the importance of precise optical alignment in future CMB polarization experiments. Throughout this work, we assume a constant wedge angle. However, the potential wobbling of the reflective HWP introduces a time-dependent variation of the wedge angle, which may have a non-negligible impact on the results of this paper. Future studies will address this effect by investigating the time-dependent variations in the wedge angle and their impact on the overall systematics, as well as exploring more realistic scenarios where detectors within each pair are not perfectly balanced.
}

\begin{document}
\maketitle
\flushbottom
\newpage
\section{Introduction}
\label{sec:intro}
Today several experiments are ongoing or planned to explore the polarization of the Cosmic Microwave Background (CMB)~\cite{2019ade, Abazajian_2022, 2018gualtieri, 2014dober, 2021addamo,10.1117/12.926581,Austermann_2012,2003SPIE.4843..284K}.
CMB polarization can be decomposed in a symmetrical component ($E$ modes) and anti-symmetrical component ($B$ modes) that can only be generated by primordial gravitational waves (tensor modes)~\cite{PhysRevLett.78.2058,PhysRevD.55.1830}, or by gravitational lensing effect on $E$-mode polarization~\cite{LEWIS20061}.
$E$-mode polarization has been detected by several experiments, but due to their faintness, primordial $B$ modes remain undetected. Their discovery will enable the estimation of the amplitude of tensor modes at recombination, commonly quantified by the tensor-to-scalar ratio, \( r \). Measuring \( r \) will provide crucial insights into the physics of the early Universe and help constrain inflationary models. Ground-based experiments such as POLARBEAR~\cite{10.1117/12.926354}, SPT~\cite{Pan_2023} and ACT~\cite{KOSOWSKY2003939} have successfully observed small-scale lensing $B$ modes (40 $\leq \ell \leq$ 300), which result from the gravitational lensing conversion of $E$ modes into $B$ modes. Primordial $B$ modes are expected to dominate over lensing $B$ modes only at large angular scales if \( r \leq 10^{-3} \) (reionization bump: $\ell \leq$ 10) or be at least comparable at intermediate scales (recombination bump: 10 $\leq \ell \leq$ 200). Currently, the upper bound on the tensor-to-scalar ratio is \( r \leq 0.032 \) at 95\,\% CL~\cite{Tristram_2022,PhysRevLett.127.151301}.\\ 
In this paper, we focus on the \LiteBIRD space mission. \LiteBIRD aims to conduct a comprehensive survey of the polarization of the CMB radiation with unparalleled precision. Its ultimate success hinges on achieving $\delta r<0.001$ for a fiducial model with $r=0$~\cite{litebird2023probing, Campeti_2024}. Here, $\delta r$ represents the total uncertainty on the tensor-to-scalar ratio. To distinguish between these primordial $B$ modes and foreground components, \LiteBIRD will conduct a comprehensive survey of the entire sky across 15 frequency bands ranging from 34 to $448\,\mathrm{GHz}$. With an effective polarization sensitivity of $2\,\mu \mathrm{K}$-arcmin and an angular resolution of $31\,\mathrm{arcmin}$ (at $140\,\mathrm{GHz}$), \LiteBIRD's observation strategy includes rapid polarization modulation and the stable orbital environment of a Lagrange point L2 orbit, offering unparalleled control over systematic errors, especially on large angular scales below $\ell \simeq 10$. This approach ensures \LiteBIRD's capability to detect both the reionization and recombination bumps in the $B$-mode power spectrum, enhancing confidence in identifying a primordial signal. Notably, if a ground-based or balloon-borne experiment hints at the recombination peak, \LiteBIRD can conclusively confirm the signal detection and significantly refine our understanding of inflationary physics~\cite{litebird2023probing}.\\
In this paper, we focus on the use of a rotating Half-Wave Plate (HWP) in the \LiteBIRD space mission~\cite{Johnson_2007,10.1117/12.857138,Hill_2016, Columbro_2022,2022pisano, Ghigna_2023, 10.1117/12.2312431, 10.1117/12.2312391}. The HWP modulates detector signals by rotating incoming polarization, which helps to reduce systematics and correlated $1/f$ detector noise~\cite{2009Brown,2016Bryan,2016Essinger,Sekimoto_2020, Stever_2021}. However, imperfections or non-idealities of the HWP can introduce systematic effects that need to be carefully addressed~\cite{giardiello2022, 2024Monelli, sugiyama2020, Verg_s_2021, Abitbol_2021}. In this work, we use the term {\sl wedge-like effect} to describe a systematic effect specific to reflective HWP configurations, arising when the HWP’s rotation axis is misaligned with the instrument’s boresight by a small constant angle, despite the HWP having a perfectly ideal shape. In transmissive configurations, similar systematic effects arise when the HWP surfaces are not parallel, a condition that can be similarly quantified by a wedge angle~\cite{litebird2023probing}. While our analysis focuses on the reflective configuration, the underlying formalism applies to both cases. The aim of this paper is to investigate how the wedge-like effect influences polarization measurements and to understand the systematic errors it introduces. To evaluate the wedge-like effect, we performed end-to-end simulations that track how these systematics propagate from raw time-ordered data (TOD) to final CMB maps and angular power spectra. This allowed us to determine constraints on the optical design to minimize its impact on observations. The paper is structured as follows: Section~\ref{sec:formalism} introduces the theoretical framework. Section~\ref{sec:wedge_intro} explains the wedge-like systematic and its impact on polarization measurements. Section~\ref{sec:litebird_case} presents the case of \LiteBIRD. Finally, Section~\ref{sec:conclusions} draws the conclusions.

\section{Formalism of the \LiteBIRD experiment}
\label{sec:formalism}

This section outlines the formalism applied throughout the paper.

\subsection{Sky Model}  

The sky emission \simone{is decomposed in the three Stokes parameters, expressed in spectral brightness units [W\,m$^{-2}$\,sr$^{-1}$\,Hz$^{-1}$]} and it's here approximated \simone{in the limit of small variations} by:  
\begin{align}
I(\nu, \hat{n}) &= I_0(\nu) + \left.\frac{\partial B(\nu, T)}{\partial T}\right|_{T_0} \Delta T_{CMB}(\hat{n}) \label{eq:I} \\
Q(\nu, \hat{n}) &= \left.\frac{\partial B(\nu, T)}{\partial T}\right|_{T_0} \Delta Q_{CMB}(\hat{n}) \label{eq:Q} \\
U(\nu, \hat{n}) &= \left.\frac{\partial B(\nu, T)}{\partial T}\right|_{T_0} \Delta U_{CMB}(\hat{n}) \label{eq:U}
\end{align}
\simone{where $\Delta T_{CMB}$, $\Delta Q_{CMB}$ and $\Delta U_{CMB}$ are the corresponding fluctuations
in units of K$_\text{CMB}$.} 
Equation~\ref{eq:I} expresses the total intensity, \( I(\nu, \hat{n}) \), as the sum of the CMB monopole term, \( I_0(\nu) \), and the CMB anisotropies modeled using the derivative of the black-body spectrum \( B(\nu, T) \) with respect to temperature at \( T_0 = (2.72548 \pm 0.00057 )\)\,K~\cite{Fixsen_2009}. Similarly, equations~\ref{eq:Q} and~\ref{eq:U} describe polarized sky emission, where a monopole term is not expected.
\simone{In this study we adopt a deliberately simplified sky model (CMB only) and do not include the dipole, instrumental noise, or astrophysical foregrounds. Our goal is to isolate and characterize the impact induced by a wedge-like effect, and to trace how this systematic propagates from TOD to maps and power spectra. Including noise and foregrounds would complicate the interpretation of small HWP-synchronous signals and could mask the pure imprint of the wedge in the plots presented here.
While this simplification prevents a direct assessment of the impact on a full mission analysis (which would require component separation and realistic noise properties), it allows a clean validation of the pipeline and an unambiguous identification of the wedge spectral signature, such as harmonics of the HWP rotation frequency (see Section \ref{sec:wedged_signal}) and a lensing-like contribution to the $BB$ spectra (See Section \ref{sec:spectra}). 
We explicitly acknowledge this limitation and plan to extend the analysis to include realistic noise and foreground realizations, and to jointly estimate the wedge parameter $\alpha$, in future work.
In a realistic context, the presence of noise and polarized foregrounds could, in principle, introduce additional degeneracies with the systematic bias induced by the wedge; however, investigating these effects is beyond the scope of this proof-of-concept study.}

\subsection{The Scanning Strategy}
\begin{figure}[t]
    \centering
\includegraphics[width=0.7\textwidth]{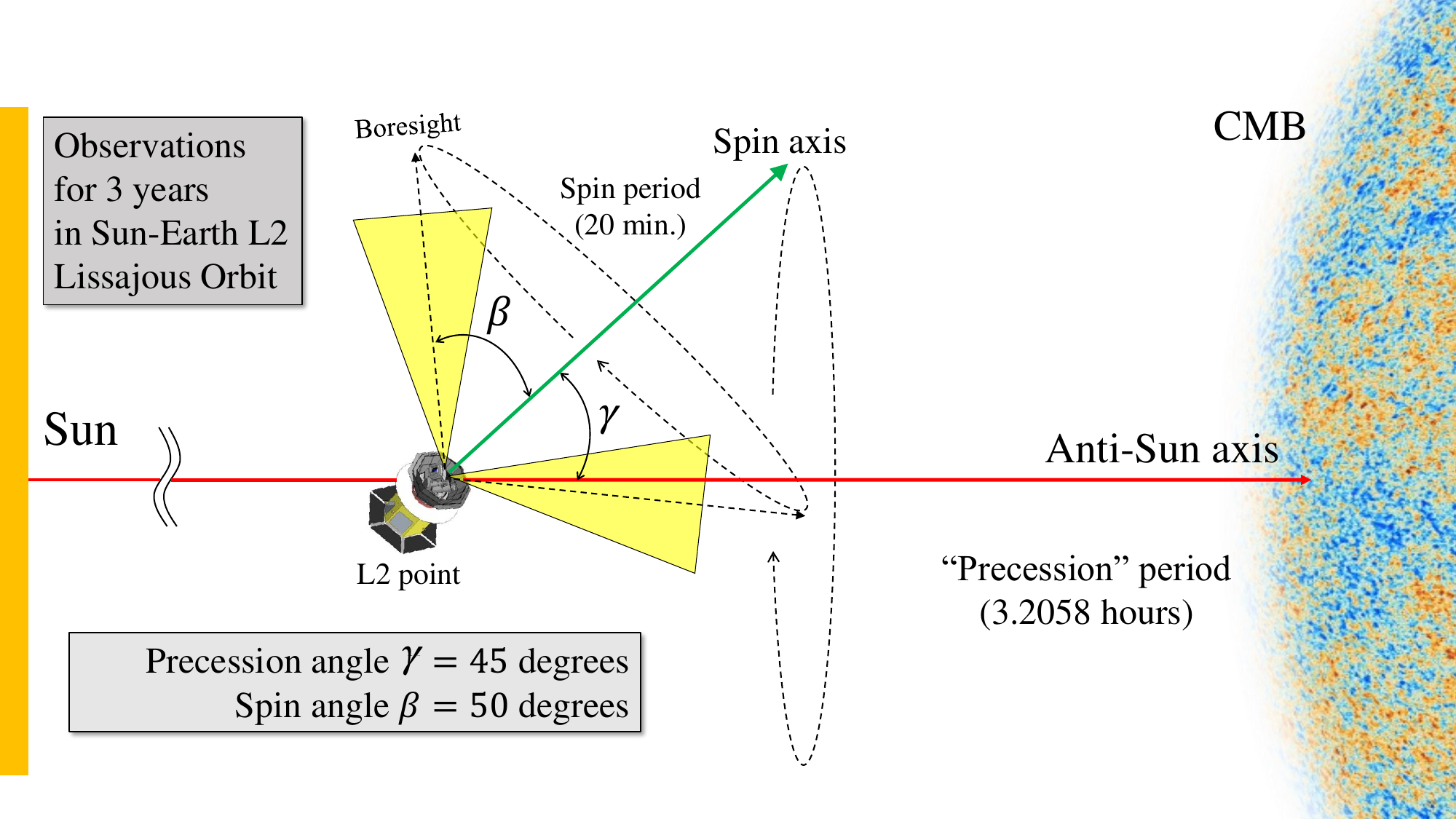}
\caption{Schematic of the observational parameters. The telescope boresight is at an angle $\beta=50^{\circ}$ to the spin axis and rotates at a rate of 0.05\,rpm. The spin axis is rotated around the anti-Sun direction through precession with an angle $\gamma=45^{\circ}$ in 3.2058\,h. The anti-Sun axis rotates around the Sun in one year. With a combination of the three motions, the boresight can cover the entire sky in half a year~\cite{litebird2023probing}. This figure is taken from Figure 37 of~\cite{litebird2023probing} and is distributed under the Creative Commons Attribution License.}
    \label{fig:scan}
\end{figure}

The scanning strategy of \LiteBIRD is designed to allow full sky coverage while ensuring uniform sampling and minimizing systematic effects. The mission employs a combination of spin and precession motions to achieve an efficient full-sky survey, with parameters carefully chosen to optimize the distribution of hits across the sky and the polarization modulation~\cite{litebird2023probing}.
\begin{table}[t]
    \centering
     \resizebox{\textwidth}{!}{ 
        \begin{tabular}{ccccc ccc c}
            \toprule
            \begin{tabular}[c]{@{}c@{}}Precession angle\\ $\gamma$\,[deg.]\end{tabular} & \begin{tabular}[c]{@{}c@{}}Spin angle\\ $\beta$\,[deg.]\end{tabular} & Precession rate\,[min.] & Spin rate\,[rpm] & \multicolumn{3}{c}{HWP revolution rate\,[rpm]} & Sampling rate\,[Hz] \\ 
            \cmidrule(lr){5-7}
            & & & & LFT & MFT & HFT & \\ 
            \midrule
            45 & 50 & 192.348 & 0.05 & 46 & 39 & 61 & 19 \\ 
            \bottomrule
        \end{tabular}
    }
    \caption{Baseline observation strategy and parameters related to the sampling rate~\cite{litebird2023probing}.}
    \label{tab:imo}
\end{table}
\noindent
Table~\ref{tab:imo} lists the primary baseline parameters common to the three telescopes: Low-, Mid- and High- Frequency Telescopes (LFT, MFT and HFT)~\cite{litebird2023probing,10.1117/1.JATIS.9.2.024003,10.1117/12.2629271}. \simone{This configuration corresponds to the \LiteBIRD baseline design, which is being revised as part of the ongoing redesign process.} Parameters related to the observational scan strategy are further illustrated in the schematics shown in Figure~\ref{fig:scan}. The scan strategy parameters, including $\gamma$, $\beta$, and the spin rate, are optimized to achieve the most uniform distributions possible for both the hits over the entire sky area and the scanning directions within individual sky pixels. A spin rate below 0.05\,rpm results in non-uniform distributions for both quantities. The precession rate is determined to avoid moiré patterns using the procedure described in~\cite{Hoang_2017}. The HWP spin rates are adjusted to prevent overlap of the HWP harmonics with the science observation frequency band, as determined by the beam FWHM. The sampling rate is set to twice the Nyquist frequency of the HWP modulation.

\subsection{Formalism of the error}
\label{sec:likelihood}
As in Section 5.3.2 of~\cite{litebird2023probing}, we define a likelihood function \(L(r)\) in the multipole domain, suitable for the large sky coverage of the satellite mission. Assuming Gaussian stationary and isotropic fields, with no coupling between multipoles (a simplification when considering the Galactic plane mask), the likelihood function is given by:
\begin{equation}
\log L(r) = \sum_{\ell=\ell_{\min}}^{\ell_{\max}} \log P_{\ell}(r)
\end{equation}
where \(\ell_{\min} = 2\) and \(\ell_{\max} = 3 \times N_{\text{side}} - 1\), with $N_{\text{side}}$ being the HEALPix resolution parameter, which determines the angular resolution of the map~\cite{Gorski_2005}.
We define:
\begin{equation}
\log P_{\ell}(r) = -f_{sky} \frac{2\ell + 1}{2} \left[\frac{\hat{C}_{\ell}}{C_{\ell}} + \log C_{\ell} - \frac{2\ell - 1}{2\ell + 1} \log \hat{C}_{\ell}\right]
\end{equation}
Here, \(\hat{C}_{\ell}\) represents the measured $B$-mode power spectrum, \(C_{\ell}\) represents the modeled $B$-mode power spectrum and \(f_{sky}\) represents the fraction of sky covered~\cite{PhysRevD.77.103013}. In the simulations used in this analysis, noise and foregrounds are not considered, and the full sky is used to compute the $C_\ell$-s and the likelihood, that is exact in this case. \simone{In this work we adopt full-sky simulations to isolate the wedge-like effect. On a cut sky, the loss of spin-2 harmonic orthogonality induces $E \rightarrow B$ leakage and mode coupling \cite{Lewis_2001, de_Oliveira_Costa_2003}, generating spurious $B$-modes which could obscure the systematic effect. Since the aim of this study is to cleanly identify the wedge imprint without additional sources of contamination, we leave a realistic partial-sky analysis, including foregrounds and noise, to future work.} To estimate the potential bias of individual systematic effects on \(r\), denoted \(\Delta r\), we represent the measured $B$-mode spectrum as the sum of the following contributions (assuming no primordial $B$ modes, i.e., \(r = 0\)):
\begin{equation}
\hat{C}_{\ell} = C_{\ell}^{\text{sys}} + C_{\ell}^{\text{lens}} + {N_{\ell}}
\end{equation}
Here, \(C_{\ell}^{\text{sys}}\) is the estimated systematic effects power spectrum, \(C_{\ell}^{\text{lens}}\) is the lensing $B$-mode power spectrum, and \(N_{\ell}\) is the expected noise, which again in our case is set to zero. The modeled power spectrum is given by:
\begin{equation}
C_{\ell} = r C_{\ell}^{\text{tens}} + C_{\ell}^{\text{lens}} + {N_{\ell}} 
\end{equation}
where \(C_{\ell}^{\text{tens}}\) is the tensor mode with \(r = 1\).
The potential systematic bias \(\Delta r\) is defined as the value that maximizes the likelihood function:
\begin{equation}
\left.\frac{dL(r)}{dr}\right|_{r = \Delta r} = 0
\end{equation}
The total error on \(r\), \(\delta r\), is defined as the value covering 68\% of the area under the total likelihood function:
\begin{equation}
\frac{\int_0^{\delta r} L(r) \, dr}{\int_0^{\infty} L(r) \, dr} = 0.68
\end{equation}

\section{The Reflective HWP’s wedge-like systematic}
\label{sec:wedge_intro}
This paper focuses on an alternative reflective configuration for the \LiteBIRD Medium-to-High Frequency Telescope (MHFT) instrument, featuring a reflective HWP. \simone{Although hypothetical, this telescope setup is used as the study case. The reflective HWP configuration was adopted as it corresponds to one of the designs considered for the MHFT during the \LiteBIRD optical studies. The conclusions drawn in this work do not depend on the specific details of the assumed optical setup, as the systematic originates within the modulator itself and the HWP is the first optical element in the system; therefore, the same behaviour would be expected for alternative optical configurations, including transmissive HWPs.} The primary objective of this research is to characterize through the \LiteBIRD simulation framework (LBS)~\cite{tomasi2025simulationframeworklitebirdinstruments} a systematic effect, specifically addressing the wedge-like effect in an ideal reflective HWP, which occur when the rotation axis of the HWP is not perfectly aligned to its optical axis (defined as the normal through the centre of the HWP), but is instead tilted by a constant angle \(\alpha_\text{{wedge}}\). This misalignment induces a systematic error in the reflected beam. As the optical component rotates, the deviation from the expected beam path becomes periodic, with a characteristic frequency corresponding to the rotation rate. This results in a pointing error, where the observed pointing direction describes a circular pattern around the nominal boresight. \simone{In LBS, this effect is implemented as a common boresight pointing offset, identical for all detectors, rather than through a full optical propagation. This approach captures the geometrical consequences of wedge-like effects and is applicable to both reflective and transmissive HWP configurations. The optical chain is otherwise kept unchanged, allowing for a clean isolation of the systematic impact on the simulated data.} This boresight error must be considered and accounted for to achieve accurate observations; the presence of circles in the sky results in a distortion of the scanning pattern, and eventually in an inaccurate reconstruction of the cosmological information.
\begin{figure}[t]
\centering
\includegraphics[width=0.8\textwidth]{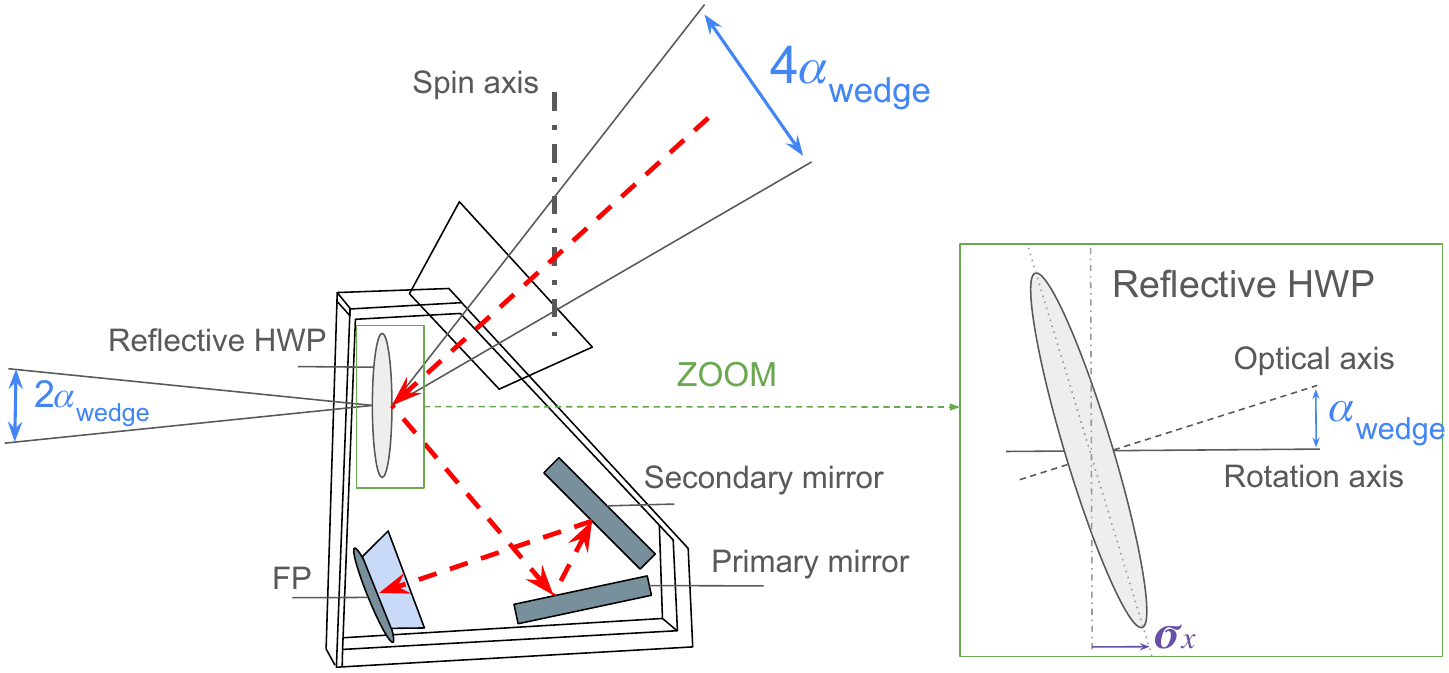}
\caption{Schematic of the optical configuration, highlighting the reflective HWP and the systematic effects introduced by the wedge angle \(\alpha_\text{wedge}\). The diagram illustrates the light propagation through the telescope and how the wedge angle modifies the optical path, introducing systematics. The inset zoom provides a closer view of the reflective HWP. The reflective HWP is assumed ideal; this misalignment mimics the effect of a physical wedge in a transmissive HWP.}
    \label{fig:wedge}
\end{figure}
\noindent
Figure~\ref{fig:wedge} shows various steps of the systematic effect induced by the wedge:
\begin{enumerate}
    \item \(\alpha_\text{wedge}\): Initial inclination of the rotation axis with respect to the normal of the HWP.
    \item \(2\alpha_\text{{wedge}}\): Resulting deviation in the reflected beam due to the inclination of the rotation axis.
    \item \(4\alpha_\text{{wedge}}\): Angular diameter of the circle in the sky resulting from the inclination and rotation.
    \item \(\sigma_x\): Displacement of the HWP edge from its nominal position along the rotation axis, induced by \(\alpha_\text{wedge}\).

\end{enumerate}
The size of the described circle depends on the inclination angle of the optical axis. To minimize this effect, it is crucial to align the rotation axis as precisely as possible with the HWP's optical axis. Note that in the case of a transmissive HWP a similar effect can be induced, if the HWP surfaces are not parallel to each other.\\
The aim of this study is to quantify the wedge-like effect to ensure accurate cosmological results. 
To accomplish this, a code has been developed using LBS to implement the wedge angle in our measurements. Subsequently, the simulated data will be analyzed.

\subsection{Wedge-like effect on pointing}
\label{sec:point}
\begin{figure}[t]
\centering
\includegraphics[width=0.8\textwidth]{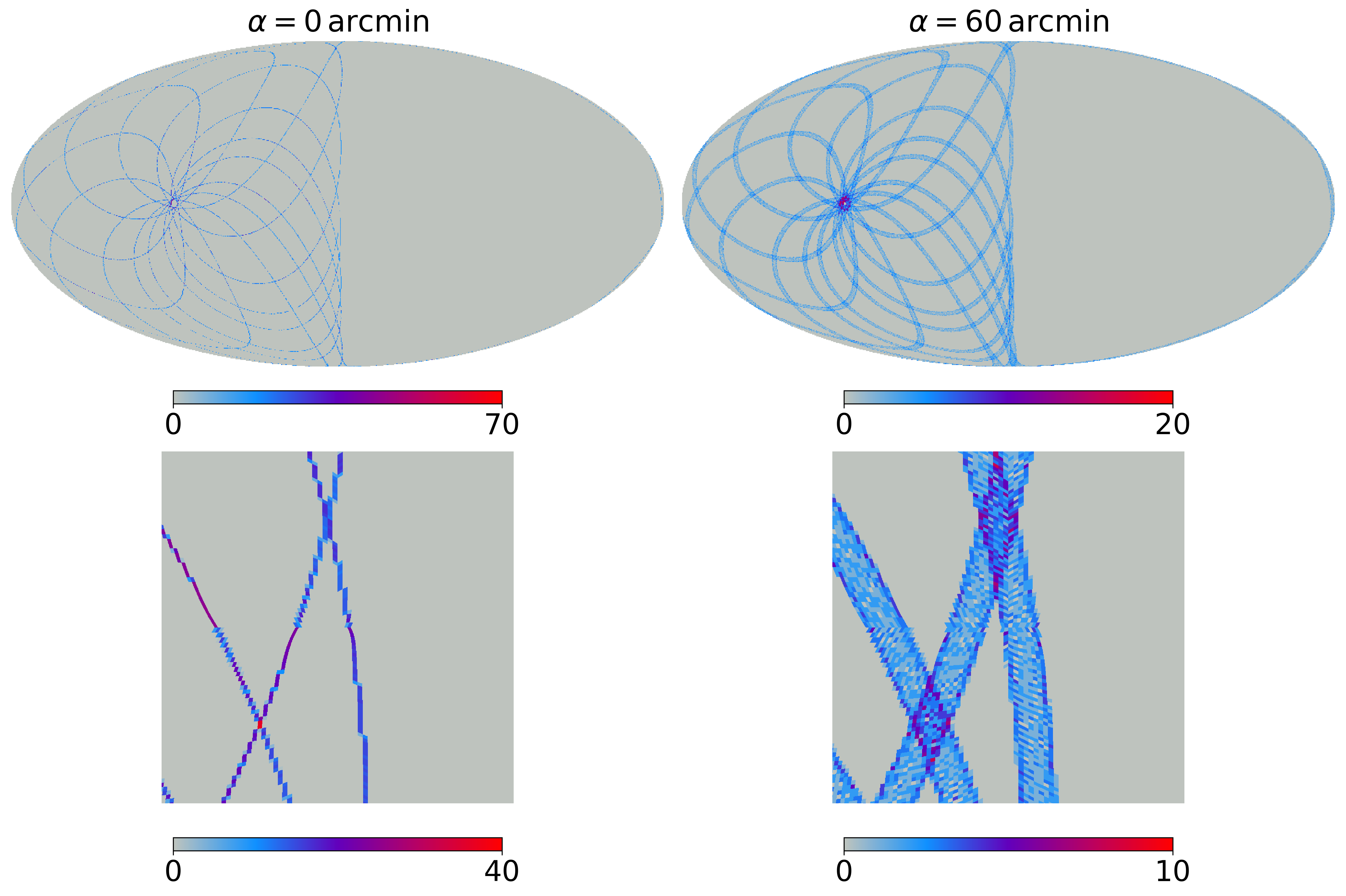}
\caption{Comparison between hit maps (mission duration: 4 hours) for different wedge angles: \(\alpha = 0\)\,arcmin (left) and \(\alpha = 60\)\,arcmin (right). Each case includes a full-sky view (top) and a zoomed-in view (bottom). The presence of a wedge angle introduces a spreading effect in the hit distribution, resulting in a broader and less concentrated scanning pattern, as evident in the right panels.}
    \label{fig:hit_map_4_hours}
\end{figure}

\begin{figure}[t]
\centering
\includegraphics[width=0.65\textwidth]{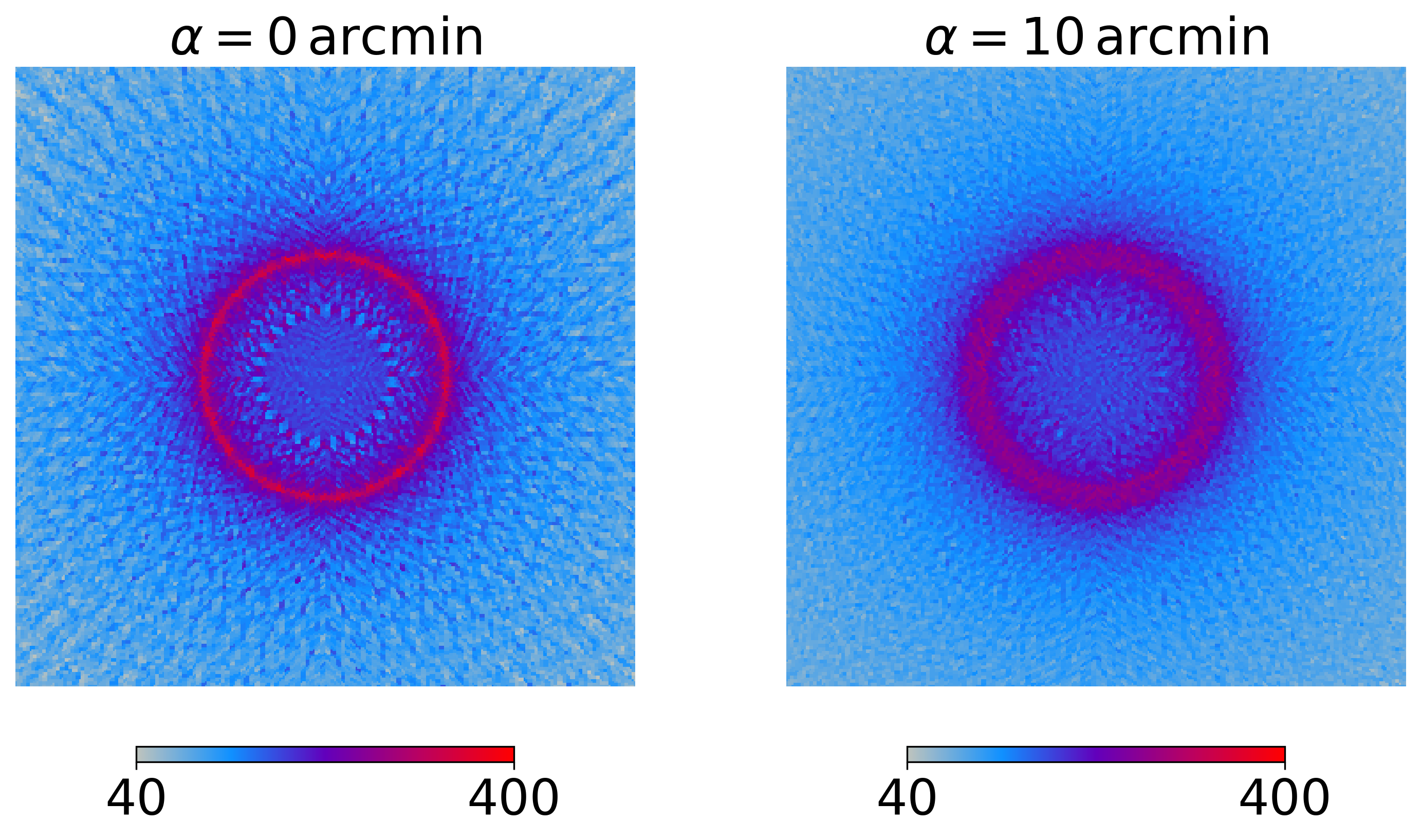}
\caption{Comparison between hit maps in zoomed-in view (mission duration: 1 year) for different wedge angles: \(\alpha = 0\)\,arcmin (left) and \(\alpha = 10\)\,arcmin (right). The maps focus on one of the most observed regions, \simone{the North Ecliptic Pole, centered at $(\mathrm{lon}, \mathrm{lat}) = (0^{\circ}, 90^{\circ})$, covering a sky patch of approximately $20^{\circ} \times 20^{\circ}$.} To facilitate a direct comparison, the same colour scale has been applied to both maps.}
    \label{fig:comp_hit_map}
\end{figure}

\begin{table}[t]
\centering
\begin{tabular}{l|l}
\hline
\textbf{Parameter} & \textbf{Value} \\
\hline
Mission duration (hours) & 4 \\
$N_{\text{side}}$  & 256 \\
Number of detectors & 1 \\
$\nu_{HWP}$ & 0.65\,Hz \\
$\nu_{sampling}$ & 19\,Hz \\
Precession angle & $45^\circ$ \\
Precession period & 192\,min \\
Spin rate & 0.05\,rpm \\
Spin angle & $50^\circ$ \\
\hline
\end{tabular}
\caption{Summary of the key simulation parameters for the analysis in Section~\ref{sec:point}.}
\label{tab:par1}
\end{table}
We begin the analysis by showing how the wedge angle affects the telescope pointing. \simone{Figure~\ref{fig:hit_map_4_hours} reports the hit maps from simulations based on the parameters in Table~\ref{tab:par1}. As the wedge angle increases, the pointing is redistributed following the circular pattern induced by the misalignment, producing a broader spread on the sky. This effect reduces the coverage uniformity, especially in short simulations where several pixels receive fewer observations. The impact is most pronounced at small angular scales, where even modest pointing errors matter.
To assess the long-term behaviour, we extended the simulation to one year at a resolution of $N_{\text{side}}=1024$. Figure~\ref{fig:comp_hit_map} shows that the presence of the wedge leads to a systematically lower hit count, reflecting the altered scanning distribution.}

\subsection{Signal measured with a wedge-angled HWP}
\label{sec:wedged_signal}
The use of a rotating HWP in combination with polarization sensitive detectors and detector averaging allows us to extract a tiny polarization fraction from the CMB intensity signal. 
The rotating HWP modulates the polarized fraction of the signal at a specific angular frequency ($4\, \omega_{HWP}$). Assuming the CMB signal has an intensity component $I$ and two polarization components $Q$ and $U$, the signal received by the detector $S(t)$, with the HWP rotating at an angular frequency $\omega_{HWP}$, can be written as:
\begin{equation}
S(t)=I+Q \cos \left(4 \omega_{\mathrm{HWP}} t-2 \phi\right)+U \sin \left(4 \omega_{\mathrm{HWP}} t-2 \phi\right)
\end{equation}
where $I$ is the total unpolarized intensity, $Q$ and $U$ are the Stokes parameters representing linear polarization, and $\phi$ is the polarization angle
of the incoming radiation with respect to the instrument's reference axis. 
A map-making algorithm can solve the $I$, $Q$, and $U$ Stokes parameters
as a function of observed sky direction, combining different observations 
by the same detector pointing in the same direction at different times, 
and combining observations by different detectors. 
The map-making effectively applies a demodulation on the single detector time-streams, 
isolating the polarization fraction. \simone{In our case, the maps are produced using the binned map-making module of LBS. Since the simulations are noise-free, the Stokes parameters are obtained by simply averaging all samples falling into each HEALPix pixel, without applying additional filtering or destriping. A wedge angle, if not included in the pointing reconstruction, modifies the true pointing as described in Section~\ref{sec:point} and produces a spurious polarization signal.}
Although highly precise optical and mechanical calibration systems exist, there is no guarantee that the wedge angle will remain consistent during flight, as variations in stress and temperature could lead to deviations.
In fact, an HWP synchronous signal generated by the pointing variation is interpreted by the map-making procedure as a polarized modulated signal. This leads to leakage from intensity into polarization, as well as mixing between the $Q$ and $U$ Stokes parameters~\cite{carretti2024cmbpolarizationmeasurements, PhysRevD.67.043004, Giardiello_2023}. In principle, the difference between the signals of two detectors with orthogonal sensitivities should cancel out the intensity $I$, leaving only the polarized component.
\begin{equation}
S_A(t) - S_B(t) = 2Q \cos (4 \omega_{\mathrm{HWP}} t - 2\phi) + 2U \sin (4 \omega_{\mathrm{HWP}} t - 2\phi)
\end{equation}
Importantly, the wedge does not induce a pointing mismatch between detectors in an orthogonal pair: both detectors view the same sky patch through the same half-wave plate, including its wedge systematic. Consequently, for perfectly matching detectors, in terms of optical response, spectral response, gain, time constant, etc., the difference of their TOD effectively cancels the intensity $I$ component, isolating only the polarized signal. This explains the absence of any significant $I \rightarrow P$ leakage in our simulations, see Section \ref{sec:maps}.
\section{Simulated Data Analysis in the \LiteBIRD case}
\label{sec:litebird_case}
\subsection{Wedge-like effect on TOD}
\label{sec:TOD}
\begin{figure}[t]
\centering
\includegraphics[width=0.7\textwidth]{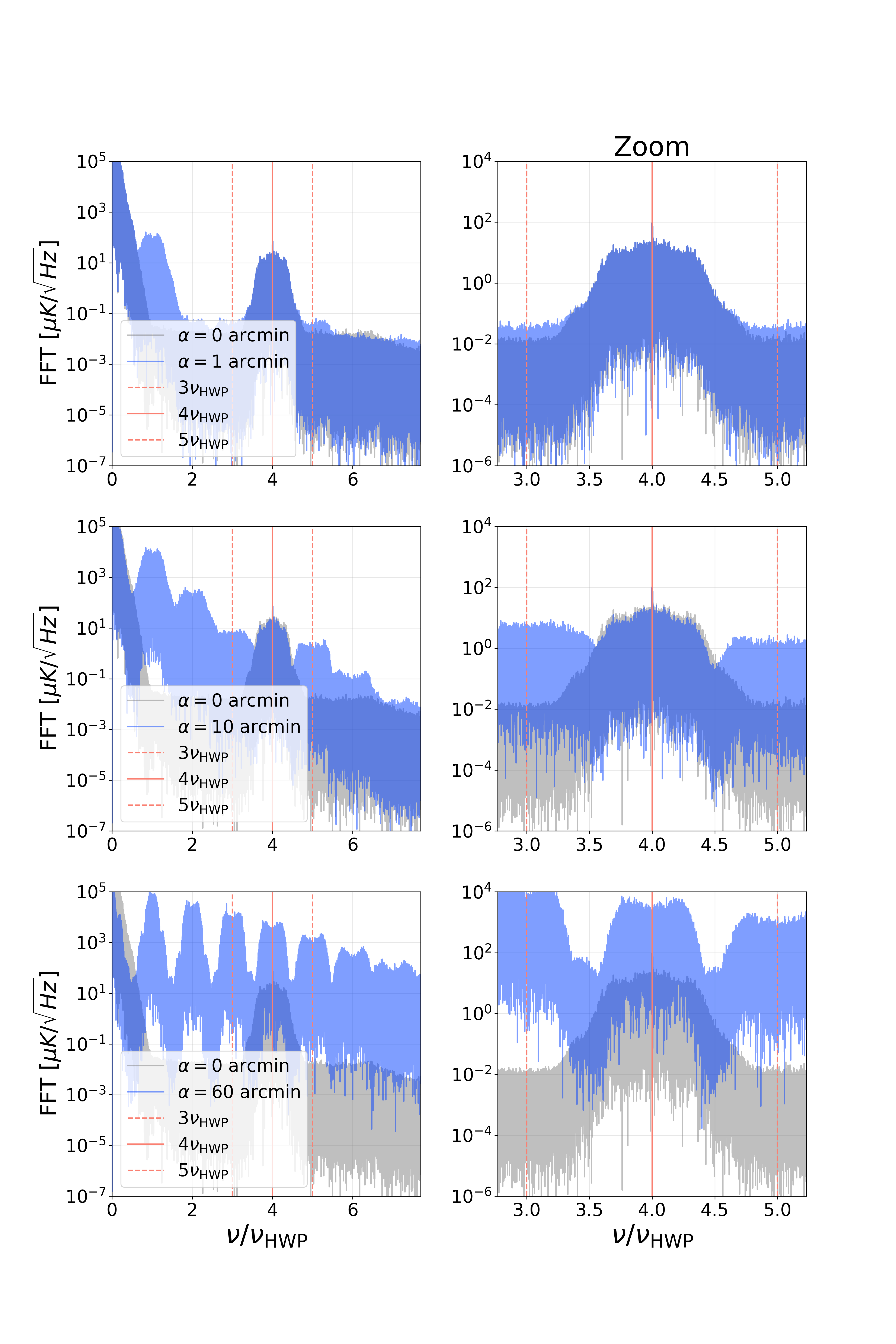}
\caption{Periodograms of the simulated timeline for different wedge angles \(\alpha\). The three rows correspond to \(\alpha = 1\)\,arcmin (first), \(\alpha = 10\)\,arcmin (second), and \(\alpha = 60\)\,arcmin (third), compared to the ideal case \(\alpha = 0\)\,arcmin (in gray). The right columns provide a zoomed-in view around \(4\,\nu_{\text{HWP}}\), highlighting the growth of additional power for increasing \(\alpha\) values. The dashed lines indicate the range \(3-5\,\nu_\text{{HWP}}\). As \(\alpha\) increases, additional harmonics of the HWP rotation frequency (orange solid line) become apparent due to the periodic nature of the effect. All frequencies are expressed in units of the HWP rotation frequency $\nu/\nu_\text{{HWP}}$.}
    \label{fig:per1}
\end{figure}

In this section, we will study how the introduction of a wedge angle impacts the TOD. This will be done through simulations with a set of different wedge angles $\alpha$. Throughout the entire section, the simulations will be carried out with the parameters shown in Table~\ref{tab:par3}, with the only modifications being the number of detectors set to 1 and the mission duration limited to 1 day. \simone{The parameters adopted for these simulations, including the number of detectors and the simplified optical configuration, are not intended to reproduce the full \LiteBIRD mission design. They were chosen to provide a controlled framework for isolating and characterizing the wedge-induced systematics. The results presented here should therefore be interpreted as a proof-of-concept demonstration of the effect rather than a mission-level forecast.}
\noindent
Since we are working with very small pointing errors comparable to the pixel size:
\begin{equation}
    \text{pixel\_size} = \sqrt{4\pi / 12{N_{\text{side}}}^2}=3.44\,\text{arcmin,}
\end{equation}
\simone{we enable in LBS the bilinear interpolation option, which evaluates the sky signal ($I,Q,U$) at each pointing direction from the four nearest HEALPix pixels. This provides a smoother and more accurate sampling of the input sky along the scanning trajectory, essential when the wedge induces sub-pixel pointing offsets.} The wedge-like effect causes the detector signal to deviate from the systematic-free case. However, the signal continues to oscillate around the mean value, which remains zero. \simone{To better characterize this behaviour, we move to the frequency domain and compute the periodograms of the simulated timelines for wedge angles \(\alpha=[0, 1, 10, 60]\)\,arcmin. Figure~\ref{fig:per1} shows the resulting spectra, which reveal the appearance of harmonics at multiples of the HWP rotation frequency ($\nu_{HWP}=0.65\,\text{Hz}$), with amplitudes increasing with \(\alpha\), consistent with the periodic nature of the effect. Moreover, additional power appears in the \(3 - 5\, \nu_{HWP}\) region, overlapping with the frequency band where the HWP-induced polarization modulation occurs. This overlap indicates that the wedge contamination may interfere with the recovery of the polarized signal.}
To verify consistency, these results have also been reproduced using input maps with a higher resolution ($N_{\text{side}} = 2048$). However, we ultimately chose to retain $N_{\text{side}} = 1024$, as the higher resolution turned out to be computationally demanding and excessively time-consuming.
Note that the observed systematic effects are driven by the wedge-like effect and arise independently of the specific structure of the input signal.
\subsection{Wedge-like effect on map-making}
\label{sec:maps}
\begin{figure}[t]
\centering
\includegraphics[width=0.7\textwidth]{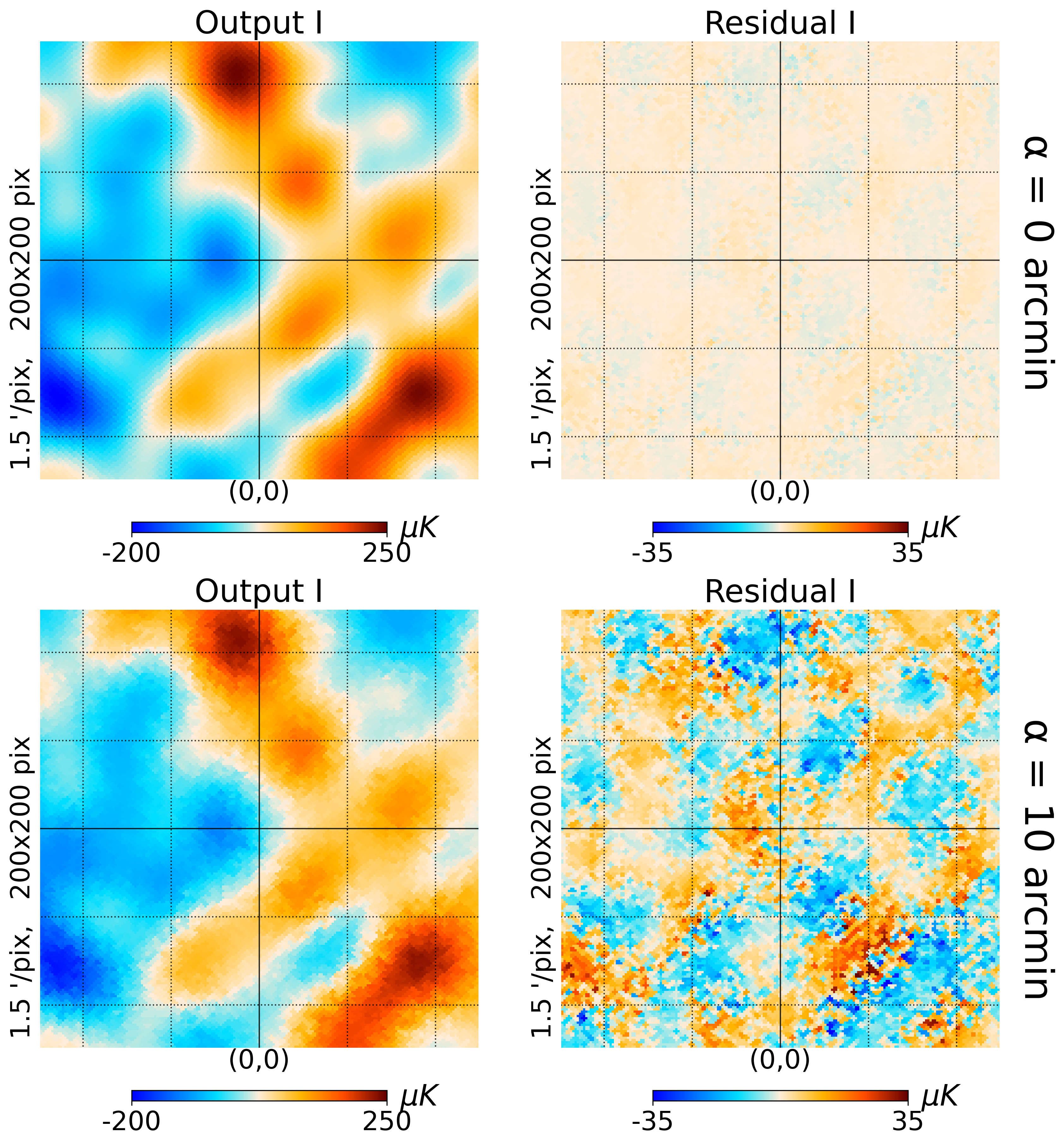}
\caption{Output and residual maps for the $I$ component in a zoomed-in view, comparing \(\alpha = [0, 10]\)\,arcmin. The left column shows the output maps, while the right column displays the residuals (Output $-$ Input). Note that, for \(\alpha = 0\)\,arcmin, the residual is not strictly zero due to the use of linear interpolation in the TOD. Simulations without interpolation yield negligible residuals (of order $10^{-12}$) for the nominal scanning strategy. However, interpolation is required to simulate the effect of small wedge angles. \simone{The maps correspond to a $5^{\circ}\!\times\!5^{\circ}$ patch centered at Ecliptic coordinates $(lon,lat)=(0^{\circ},0^{\circ})$.}}
    \label{fig:T_maps_gnom}
\end{figure}

\begin{figure}[t]
\centering
\includegraphics[width=0.7\textwidth]{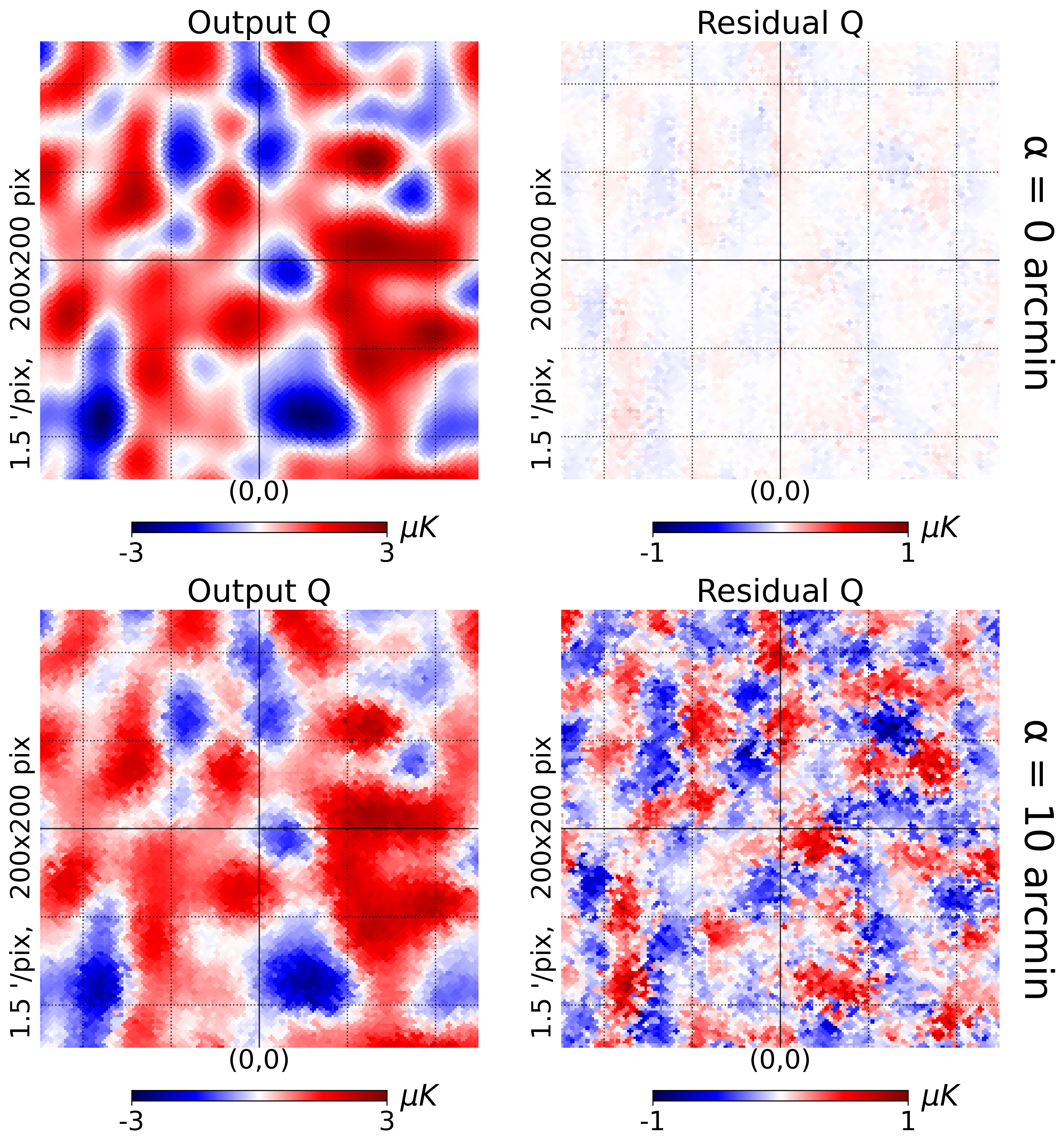}
\caption{Output and residual maps for the $Q$ component in a zoomed-in view, comparing \(\alpha = [0, 10]\)\,arcmin. The left column shows the output maps, while the right column displays the residuals (Output $-$ Input). Note that, for \(\alpha = 0\)\,arcmin, the residual is not strictly zero due to the use of linear interpolation in the TOD. Simulations without interpolation yield negligible residuals (of order $10^{-14}$) for the nominal scanning strategy. However, interpolation is required to simulate the effect of small wedge angles. \simone{The maps correspond to a $5^{\circ}\!\times\!5^{\circ}$ patch centered at Ecliptic coordinates $(lon,lat)=(0^{\circ},0^{\circ})$.}}
    \label{fig:Q_maps_gnom}
\end{figure}

\begin{figure}[t]
\centering
\includegraphics[width=0.7\textwidth]{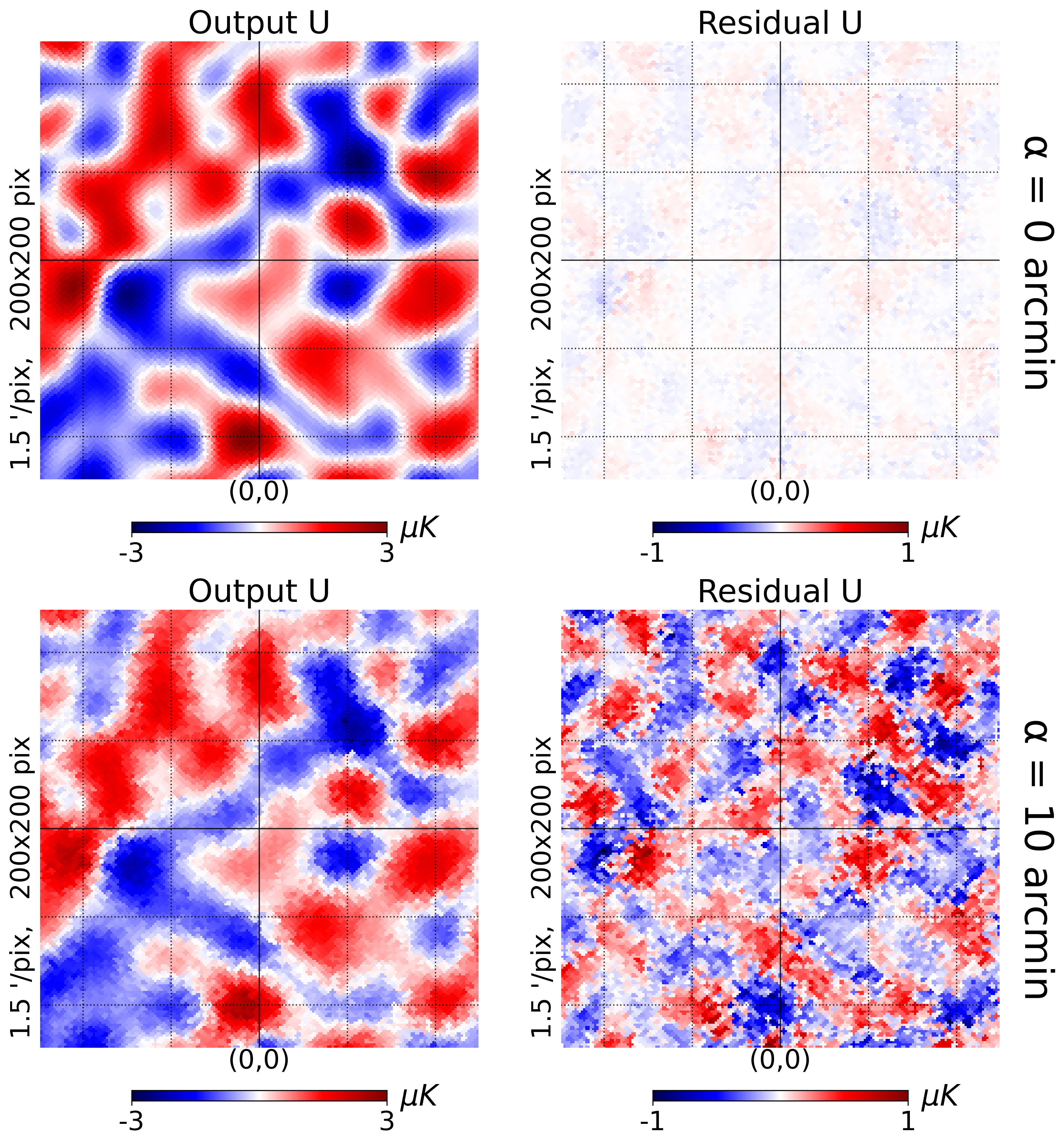}
\caption{Output and residual maps for the $U$ component in a zoomed-in view, comparing \(\alpha = [0, 10]\)\,arcmin. The left column shows the output maps, while the right column displays the residuals (Output $-$ Input). Note that, for \(\alpha = 0\)\,arcmin, the residual is not strictly zero due to the use of linear interpolation in the TOD. Simulations without interpolation yield negligible residuals (of order $10^{-14}$) for the nominal scanning strategy. However, interpolation is required to simulate the effect of small wedge angles. \simone{The maps correspond to a $5^{\circ}\!\times\!5^{\circ}$ patch centered at Ecliptic coordinates $(lon,lat)=(0^{\circ},0^{\circ})$.}}
    \label{fig:U_maps_gnom}
\end{figure}

\begin{figure}[t]
\centering
\includegraphics[width=0.7\textwidth]{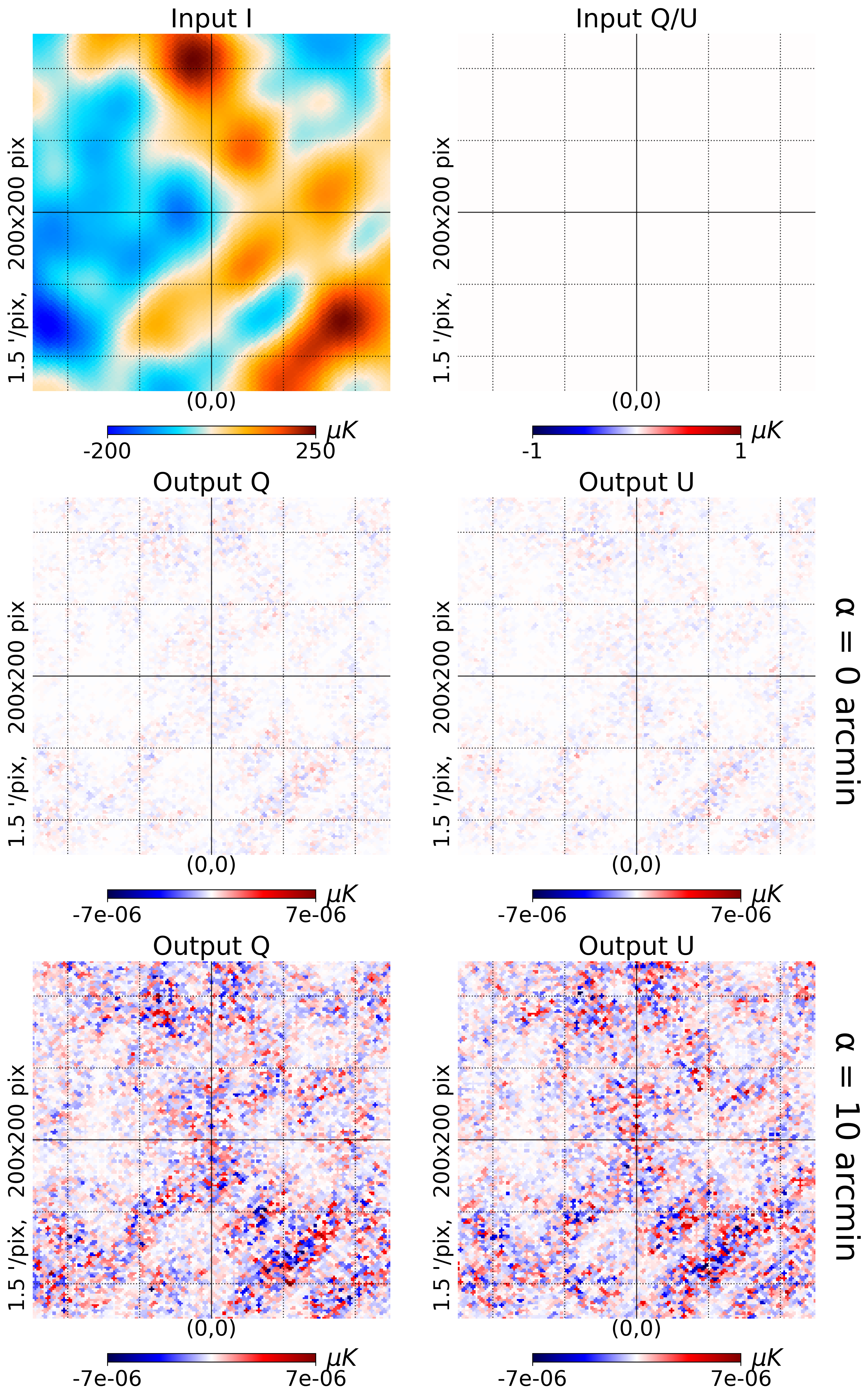}
\caption{Input and output maps for the $I$, $Q$, and $U$ components in a zoomed-in view. The top row shows the input intensity ($I$) and polarization ($Q$/$U$), where the $Q$/$U$ input maps are set to zero. The middle and bottom rows display the output $Q$ and $U$ maps for \(\alpha=[0, 10]\) arcmin. Note that, for \(\alpha = 0\)\,arcmin, the residual is not strictly zero due to the use of linear interpolation in the TOD. Simulations without interpolation yield negligible residuals (of order $10^{-14}$) for the nominal scanning strategy. However, interpolation is required to simulate the effect of small wedge angles. \simone{The maps correspond to a $5^{\circ}\!\times\!5^{\circ}$ patch centered at Ecliptic coordinates $(lon,lat)=(0^{\circ},0^{\circ})$.}}
    \label{fig:leak_maps_gnom}
\end{figure}

With the goal of obtaining full-sky scans, we run multiple simulations for different $\alpha$ values, and then proceed with performing the map-making of CMB only signals using LBS, so we are not including dipole and foregrounds contribution to our TOD. Throughout the following sections, the simulations will be carried out with the parameters shown in Table~\ref{tab:par3}.
\begin{table}[t]
\centering
\begin{tabular}{l|l}
\hline
\textbf{Parameter} & \textbf{Value} \\
\hline
Mission duration (days) & 365 \\
$N_{\text{side}}$  & 1024 \\
Gaussian beam, FWHM & 30.8\,arcmin \\
Number of detectors & 2 (orthogonal pair) \\
$\nu_{HWP}$ & 0.65\,Hz \\
$\nu_{sampling}$ & 19\,Hz \\
Interpolation & linear \\
Precession angle & $45^\circ$ \\
Precession period & 192\,min \\
Spin rate & 0.05\,rpm \\
Spin angle & $50^\circ$ \\
\hline
\end{tabular}
\caption{Summary of the key simulation parameters for the analysis in Section~\ref{sec:litebird_case}. In Section~\ref{sec:TOD}, the simulation was performed with 1 detector and a mission duration of 1 day. In Section~\ref{sec:four}, 2, 4 and 6 detectors (1, 2 and 3 orthogonal pairs) were considered. All other parameters remain unchanged.}
\label{tab:par3}
\end{table}
\noindent
It is very important to specify that we are assuming the worst-case scenario, where there is no knowledge of the presence of a wedge angle. With perfect knowledge of the wedge, the problem becomes predictable. We would easily reconstruct the pointing matrix by means of a correction in the model of the scanning strategy. For this reason, in the following map-making cases that account for systematics, the procedure involves generating a timeline based on the wedge-modified scanning path, while the map projection step is carried out using the nominal scanning strategy, i.e.,without the wedge. This is because, {\sl a priori}, we do not know whether the wedge is present at all. Also, we do not consider the possibility of treating the wedge angle as a free parameter to be retrieved {\sl a posteriori} from the collected data. \simone{Estimating $\alpha_{\mathrm{wedge}}$ from observations would require a realistic data model including noise and polarized foregrounds. Such an analysis goes beyond the proof-of-concept scope of this work, which focuses on isolating the deterministic imprint of a known wedge angle.} \simone{It should also be noted that the impact of an imperfect calibration of the wedge angle has not been explicitly addressed in this analysis. In practice, the estimated value $\alpha_{\mathrm{cal}}$ may differ from the true one $\alpha_{\mathrm{true}}$, leaving a residual misalignment $\Delta\alpha = \alpha_{\mathrm{true}} - \alpha_{\mathrm{cal}}$. Since the induced bias on $r$ does not scale linearly with $\alpha$, even a small calibration error could lead to a residual bias. A full assessment of this effect, including a realistic calibration pipeline, is left for future work, although the present results already indicate that accurate knowledge of the wedge geometry is crucial to avoid biasing the measurement of $r$.}\\
Our pipeline consists of several steps, outlined as follows:
\begin{enumerate}

\item The first step of our analysis is to generate maps using LBS containing only the CMB, with all three components: $I$, $Q$, and $U$. These are the maps that will be scanned while varying $\alpha$.

\item \simone{The second step consists of computing the residual maps. These are defined as the difference between the output maps, which include systematics, and the corresponding input maps. Although residuals could also be computed as \( M(\alpha) - M(0) \), given that we use interpolation, we found the differences between the two definitions to be negligible. Only the case \(\alpha = 0\) shows interpolation-related residuals, as the wedge contribution dominates otherwise. This definition was chosen to ensure consistency and comparability of the results across different studies like~\cite{Carralot_2025, Giardiello_2025}. Subsequently, we compute the angular power spectra for both the output and residual maps.}

\item The third step involves quantifying the total error induced by the wedge on the estimation of the tensor-to-scalar ratio, \simone{defined as}
\begin{equation}
\Delta r_{\text{wedge}} = r_{\text{fit}}(\alpha \neq 0) - r_{\text{fit}}(\alpha = 0),
\end{equation}
\simone{where \( r_{\text{fit}} \) is the value that maximizes the likelihood function. Finally, we derive the wedge-pointing error requirement, i.e. the precision with which the HWP must be aligned to meet the target sensitivity on \(r\).}
\end{enumerate}
We present in Figures~\ref{fig:T_maps_gnom},~\ref{fig:Q_maps_gnom} and~\ref{fig:U_maps_gnom} the maps obtained following the procedure just described. The figures show the zoomed-in view maps of the three components ($I$, $Q$, $U$) for different wedge angles \(\alpha\). In all three cases, we observe that as $\alpha$ increases, the residual map (right column) grows larger, exhibiting increasing contamination with $\alpha$, the power spectra of which are analyzed in Section~\ref{sec:spectra}. Notably, the $Q$ and $U$ components exhibit a higher sensitivity to the wedge-like effect, with contamination growing more rapidly with \(\alpha\) compared to the $I$ component. As mentioned in Section~\ref{sec:wedged_signal}, because our simulation uses orthogonal detector pairs, we do not expect any significant \( I \rightarrow P \) leakage. To test this, we run a simulation with the same parameters as before but without polarization input in our simulation and examine the output maps for the polarization components. The resulting output maps will quantify the effect of the intensity to polarization leakage. In Figure~\ref{fig:leak_maps_gnom}, we observe that the \( I \rightarrow P \) leakage contribution is negligible both in the absence of a wedge and in the presence of a wedge angle of \(\alpha = 10\) arcmin. It is important to note that these results hold under the optimistic assumption of perfectly matching detectors within each orthogonal pair. In practice, however, achieving such perfect balance is challenging, and detector mismatch could lead to non-negligible \( I \rightarrow P \) leakage. Future work will focus on incorporating more realistic effects, including detector mismatch, HWP wobbling, and Gaussian pointing noise.

\subsection{Wedge-like effect on the angular power spectra}
\label{sec:spectra}
\begin{figure}[t]
\centering
\includegraphics[width=0.75\textwidth]{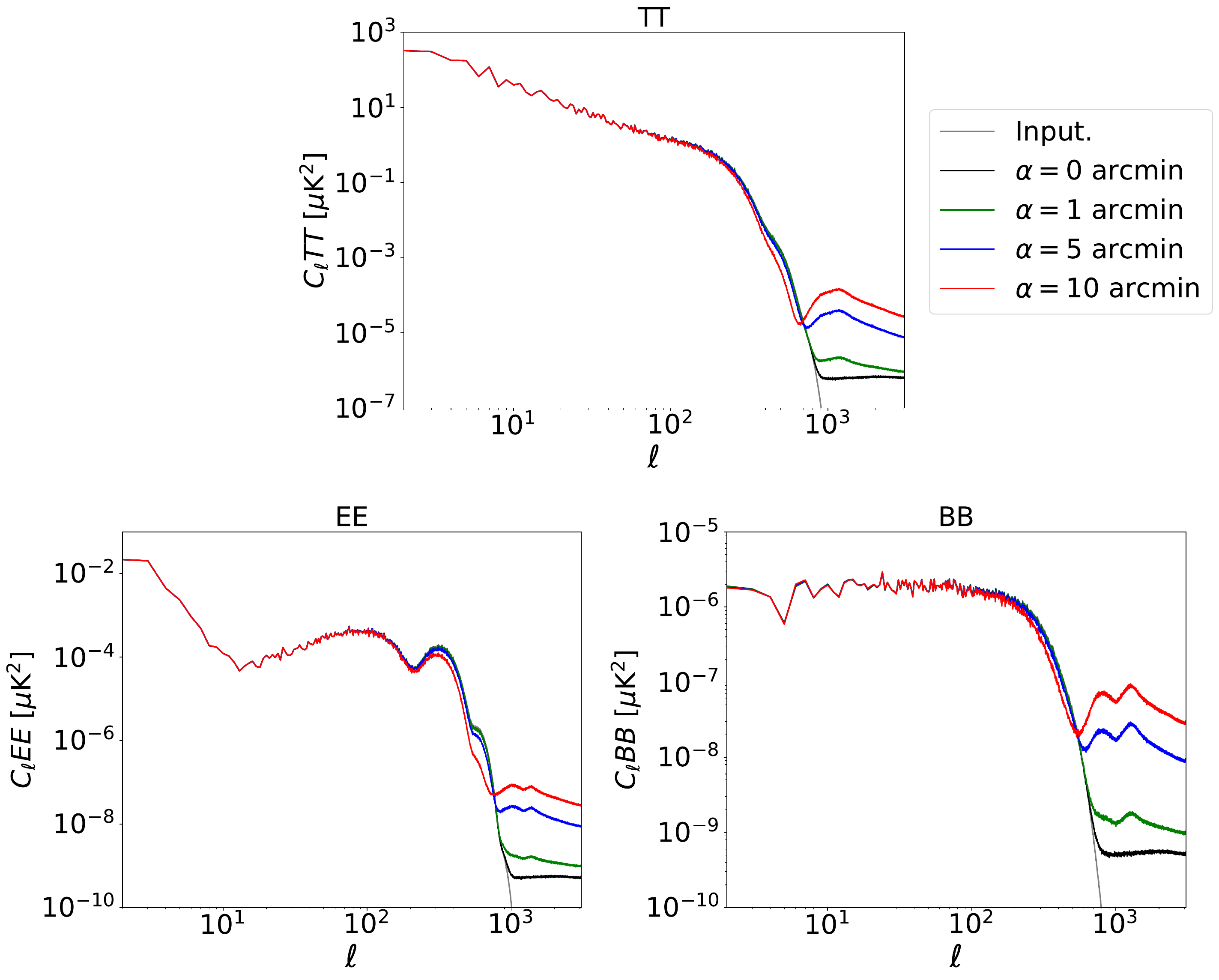}
\caption{Angular power spectra (\(C_{\ell}\)) of the output maps for different \(\alpha\) values. The panels show the $TT$, $EE$, and $BB$ power spectra. The gray line represents the input spectrum, while the colored lines correspond to different wedge angle values ranging from \(\alpha= 0\) to 10\,arcmin.}
    \label{fig:ang_spectra_output}
\end{figure}

\begin{figure}[t]
\centering
\includegraphics[width=0.75\textwidth]{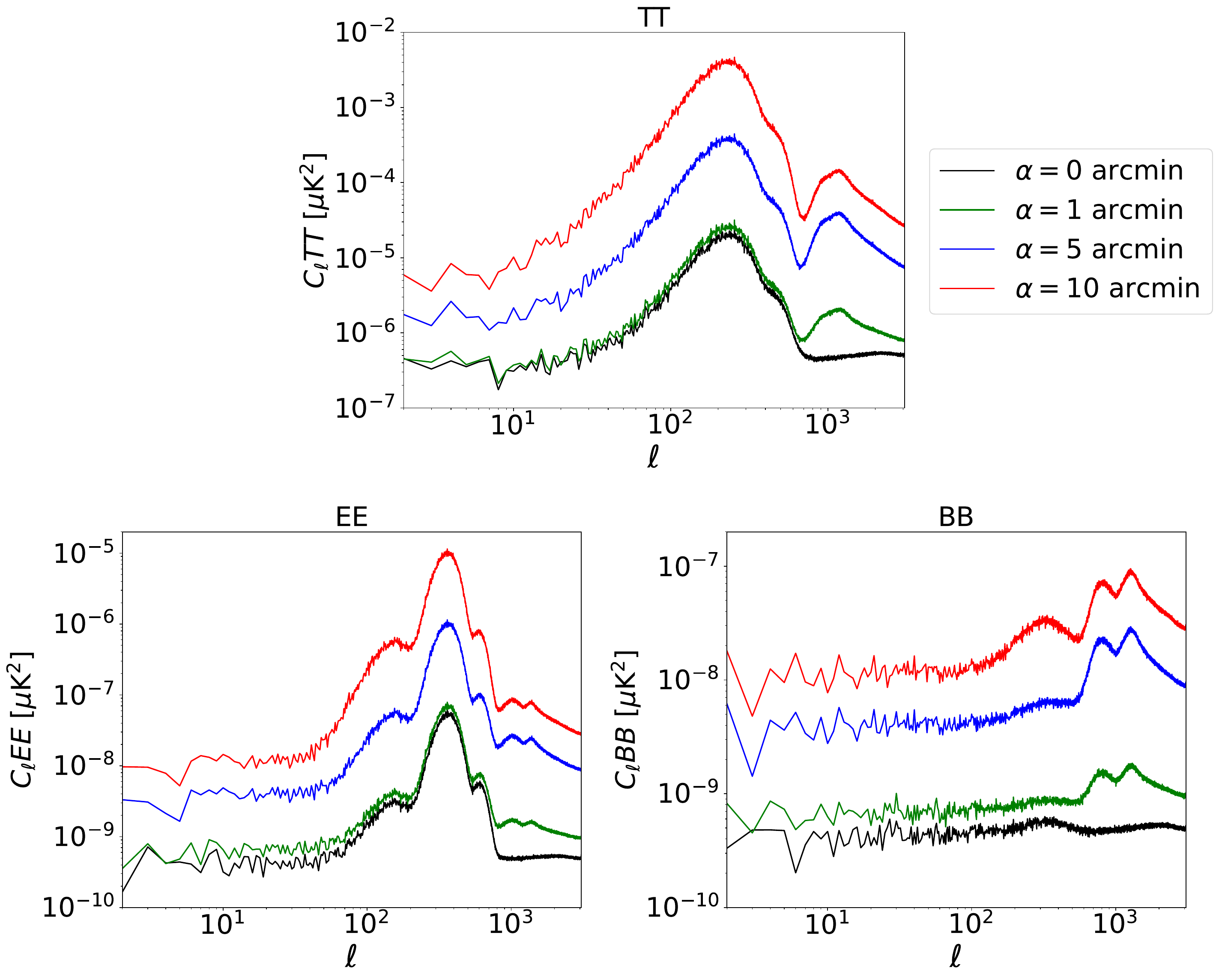}
\caption{Angular power spectra (\(C_{\ell}\)) of the residual maps for different \(\alpha\) values. The panels show the $TT$, $EE$, and $BB$ power spectra. The colored lines correspond to different wedge angle values ranging from \(\alpha= 0\) to 10\,arcmin. A small peak around \(\ell \sim 400\) is visible in both the $TT$ and $EE$ residual power spectra, even for \(\alpha = 0\). This feature is caused by the use of linear interpolation in the TOD and disappears when interpolation is disabled.}
    \label{fig:diff_ang_spectra_output}
\end{figure}

We continue the study by analyzing the angular power spectra of the maps obtained for different values of $\alpha$.\\
Figure~\ref{fig:ang_spectra_output} shows the angular power spectra of the output maps for the $TT$, $EE$, and $BB$ components. The curves represent different configurations of the angle \(\alpha\), which ranges from 0 to 10\,arcmin, with each color corresponding to a different value of \(\alpha\). We can see how the presence of a wedge angle in the HWP significantly impacts the results, making the curves deviate strongly from their expected behavior at small angular scales. Figure~\ref{fig:diff_ang_spectra_output}, shows the angular power spectra of the residual maps for the $TT$, $EE$, and $BB$ components as a function of the wedge angle $\alpha$, which ranges from 0 to 10\,arcmin. The curves represent the residual power after subtracting the ideal signals from the simulated maps, highlighting systematic effects and distortions induced by the wedge.\\
Essentially, we observe that for larger values of \(\alpha\), the deviations from the systematic-free case become more pronounced across all scales. Notably, the $BB$ residual spectrum remains almost flat, with no bump forming at low $\ell$ due to this effect. Instead, the impact is comparable to an additional white noise contamination. We note the presence of a small feature around \(\ell \sim 400\) in the $TT$ and $EE$ residual power spectra, also visible for \(\alpha = 0\). This is not related to the wedge-like effect but stems from the use of linear interpolation in the TOD. Simulations without interpolation do not exhibit this feature.


\subsection{Estimation of the tensor-to-scalar ratio}
\label{sec:dr}
\begin{figure}[t]
\centering
\includegraphics[width=0.75\textwidth]{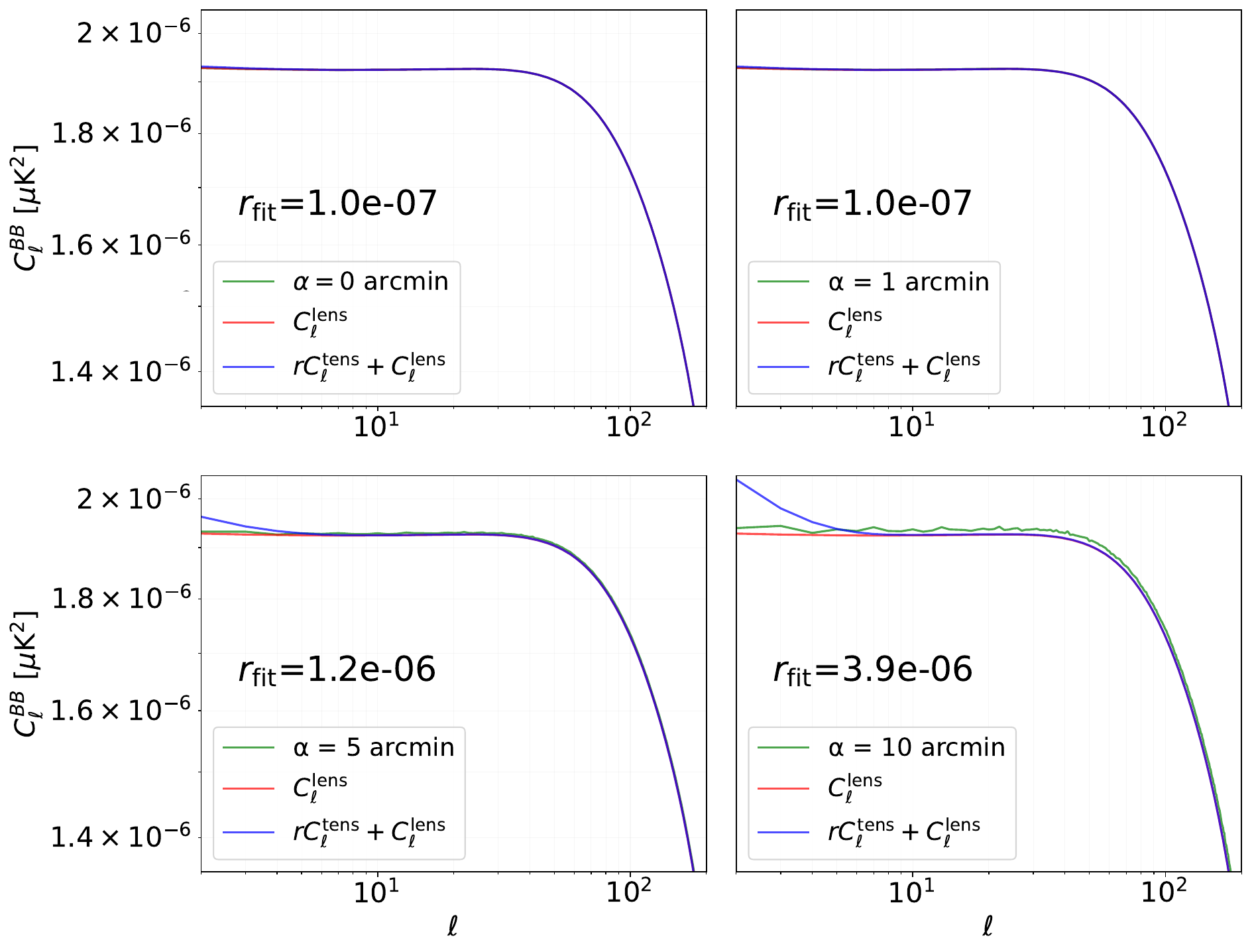}
\caption{B-mode ($BB$) power spectra of the reconstructed maps for 
$\alpha = [0, 1, 5, 10]$ arcmin. In each panel, the simulated spectrum (green) is compared with the lensing-only prediction \( C_{\ell}^{\text{lens}}\) (red) and with the fit model $C_\ell = r\,C_\ell^{\rm tens} + C_\ell^{\rm lens}$ (blue). Increasing $\alpha$ leads to an excess of power relative to the lensing case and a corresponding increase in the fitted $r_{\rm fit}$.
}
    \label{fig:quad_fit}
\end{figure}

The primary focus of our analysis is to estimate the tensor-to-scalar ratio using the procedure described in Section~\ref{sec:likelihood}. This is done through the $BB$ spectra, allowing us to determine the maximum tolerable error for the precision required in aligning the HWP, and in its intrinsic flatness. It is important to note, that \LiteBIRD's sensitivity in terms of multipoles $\ell$ ranges from 2 to 200~\cite{litebird2023probing}. Therefore, for the next stages of the analysis, we will focus exclusively on this multipole range.\\
\simone{Figure~\ref{fig:quad_fit} presents $C_\ell^{BB}$ for different wedge angles $\alpha$ in the multipole range relevant to this analysis. For each $\alpha$ we compare the simulated spectrum with the lensing-only prediction and derive $r$ assuming the model} 
\begin{equation}
        C_{\ell} = r C_{\ell}^{\text{tens}} + C_{\ell}^{\text{lens}}.
\end{equation}
\simone{As $\alpha$ increases, the reconstructed spectra exhibit a systematic excess of power relative to the lensing-only case. As a consequence, the inferred value $r_{\rm fit}$ increases with $\alpha$, with the effect becoming significant for $\alpha \geq 5$ arcmin. This indicates that the simple model above cannot absorb the systematic contribution by rescaling the primordial tensor term alone. A more detailed discussion of the limitations of this fitting approach and the resulting parameter bias is presented in Section~\ref{sec:fake}.}

\subsection{Pointing error requirements}
\label{sec:req}
\begin{figure}[t]
\centering
\includegraphics[width=0.55\textwidth]{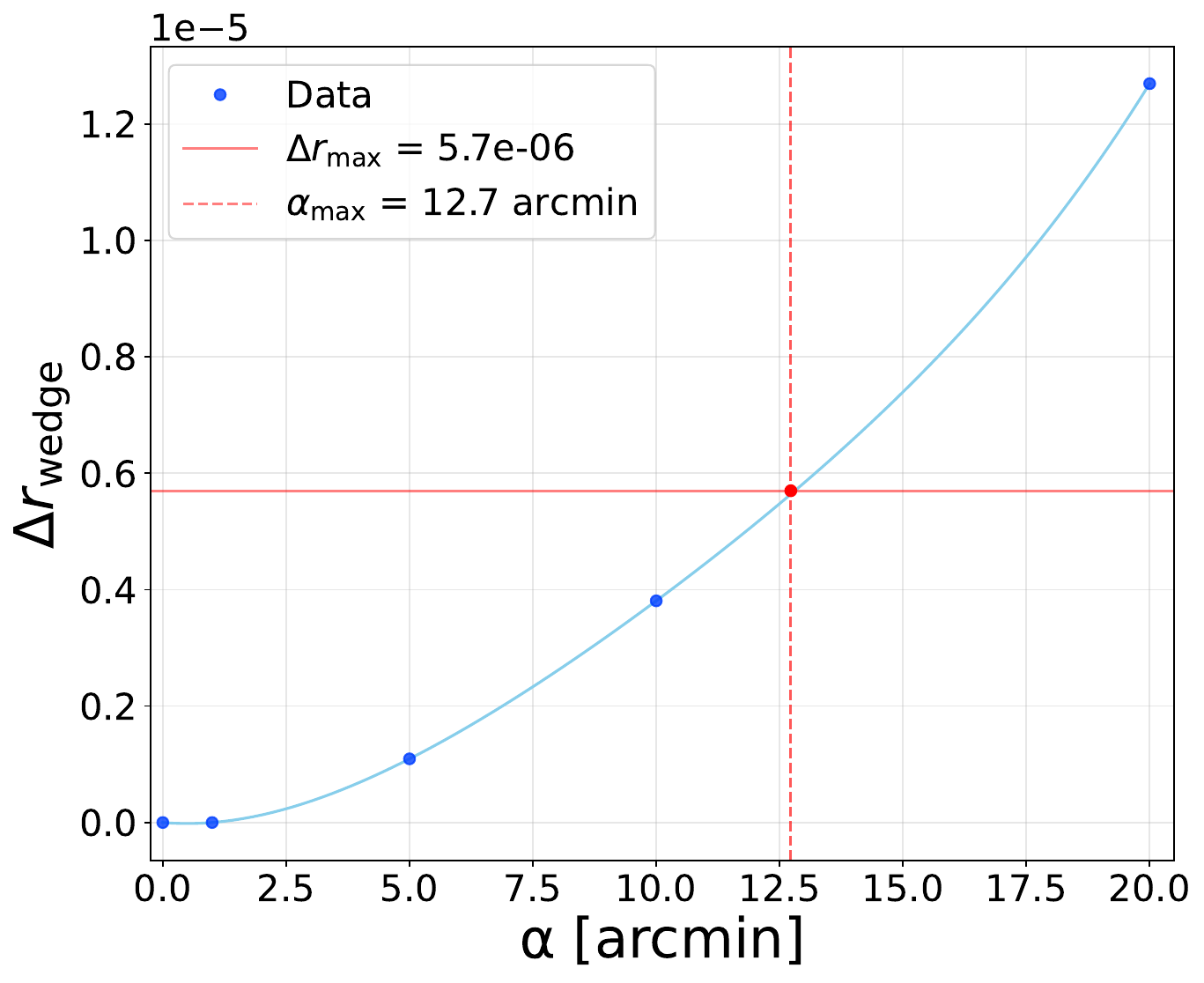}
\caption{Dependence of \simone{\(\Delta r_{\text{wedge}}\)} on the wedge angle \(\alpha\). The blue points represent simulated data, while the light blue curve is a 4th-order polynomial fit. A quadratic fit was initially tested, but it produced negative values of \simone{\(\Delta r_{\text{wedge}}\)} for some angles. The 4th-order model provides a better agreement with the data, while higher-order terms remain subdominant. The horizontal red solid line marks the maximum tolerable uncertainty \(\Delta r_{\max} = 5.7 \times 10^{-6}\), and the vertical red dashed line indicates the corresponding wedge angle limit \(\alpha_{\max} = 12.7\)\,arcmin.}
    \label{fig:dr_wedge}
\end{figure}

Let us proceed with the last step of our pipeline described in Section~\ref{sec:maps} by estimating a maximum threshold for the wedge angle $\alpha_\text{{max}}$, for which we can achieve an acceptable scientific result. This corresponds to the maximum angle for which we have a \simone{\(\Delta r_{\text{wedge}}\)} equal to 1\% of the budget assigned to all systematics for the \LiteBIRD mission, that is, \simone{\(\Delta r_\text{{max}} = 5.7 \times 10^{-6}\)}~\cite{litebird2023probing}. To do this, we study how \simone{\(\Delta r_{\text{wedge}}\)} varies as a function of \(\alpha\).\\
\simone{We report the results in Figure~\ref{fig:dr_wedge}. The results exhibit a monotonic increase of $\Delta r_{\text{wedge}}$ with $\alpha$ and we identify the maximum allowed wedge angle $\alpha_{\mathrm{max}}$ by comparing the measured $\Delta r_{\text{wedge}}$ with the requirement $\Delta r_{\mathrm{max}}$ and determining the value of $\alpha$ at which this threshold is reached via a polynomial fit. This provides a quantitative tolerance on the HWP alignment accuracy required to avoid biasing $r$ measurements. The estimated values of $\Delta r_{\text{wedge}}$} were fitted using a 4th-order polynomial. While a quadratic behavior would be expected from the linear scaling of the systematic effect at map level, a 2nd-order fit led to negative values of $\Delta r_{\text{wedge}}$ in some regions. The 4th-order model provides a more accurate fit across the explored range of $\alpha$, although higher-order terms contribute only marginally.\\
The maximum value of the wedge angle in this analysis, is given by 
\begin{equation}
\alpha_{max} = 12.7\,\text{arcmin}
\end{equation}
\simone{As an illustrative example, assuming an HWP diameter of 500 mm — a realistic order-of-magnitude value — the corresponding edge displacement from its nominal position (see Figure \ref{fig:wedge}) would be $\sigma_x \simeq 2,\text{mm}$. Such a requirement is mechanically plausible for a large cryogenic rotating element. The exact HWP dimension for LiteBIRD is still under refinement; in a reflective architecture the diameter could be even larger, which is one of the reasons why that option is generally disfavoured in the current design trade-offs.}

\subsection{Spurious lensing from the wedge-like effect}
\label{sec:fake}
\begin{figure}[t]
\centering
\includegraphics[width=0.75\textwidth]{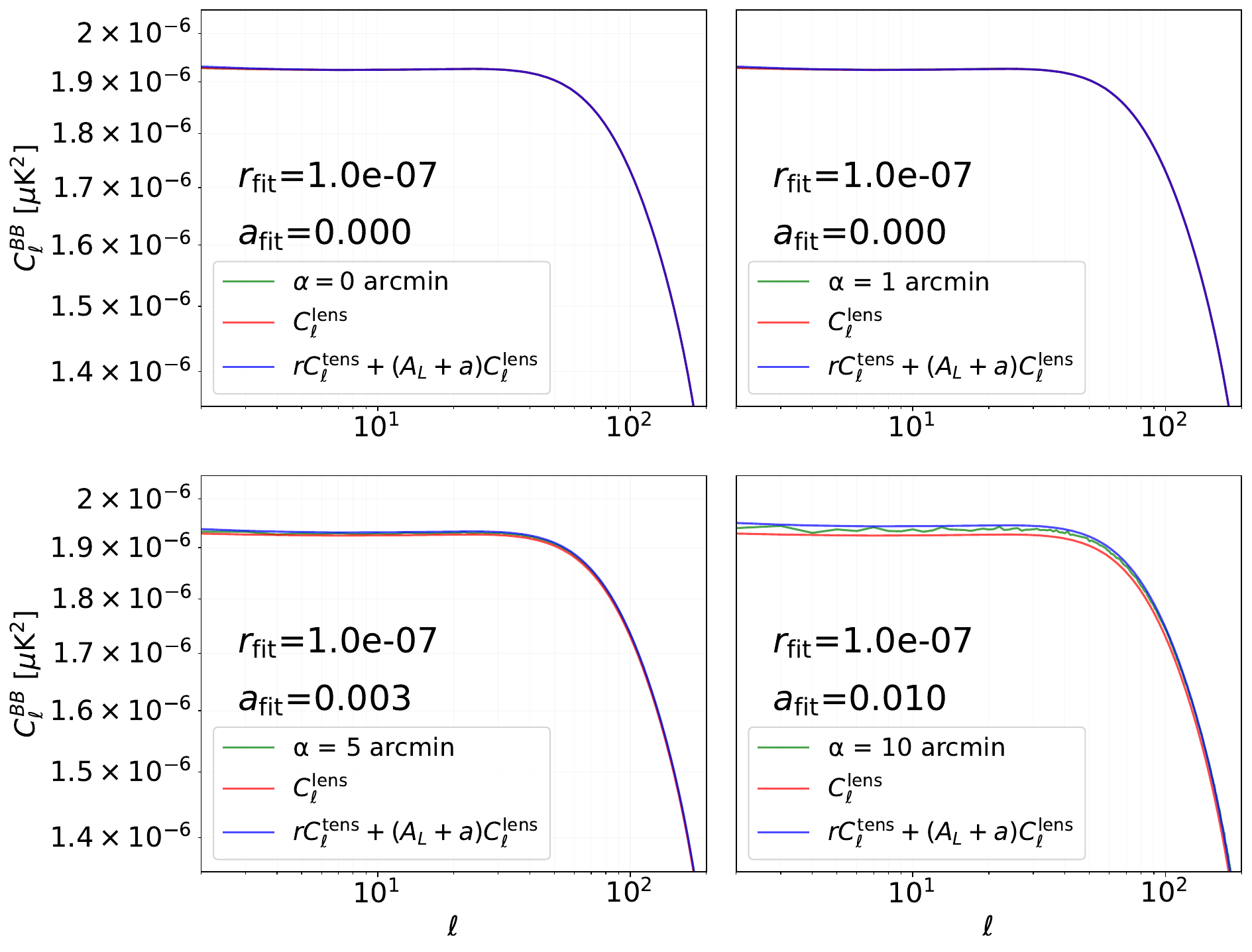}
\caption{B-mode ($BB$) power spectra of the reconstructed maps for $\alpha = [0, 1, 5, 10]$ arcmin. \simone{In each panel, the simulated spectrum (green) is compared with the lensing-only prediction \( C_{\ell}^{\text{lens}}\) (red) and with the fit model \(C_\ell = r\,C_\ell^{\rm tens} + (A_L + a)\,C_\ell^{\rm lens}\) (blue). Increasing $\alpha$ leads to an excess of power relative to the lensing case and a corresponding increase in the fitted $a_{\rm fit}$.
}}
\label{fig:quad_fit_2d}
\end{figure}

\simone{As shown in the Section~\ref{sec:dr}, the presence of a wedge angle introduces additional $B$-mode power that cannot be fully absorbed by a single-parameter fit to the primordial tensor amplitude. To capture this effect we adopt a two-parameter model of the form}
\begin{equation}
C_\ell = r\,C_\ell^{\rm tens} + (A_L + a)\,C_\ell^{\rm lens},
\end{equation}
\simone{where $r$ is the tensor-to-scalar ratio and $A_L = 1$ denotes the fiducial lensing amplitude. The parameter $a$ quantifies any residual lensing-like contribution induced by the wedge. This parametrization reflects the fact that the wedge-induced systematics produces a non-tensor component whose spectral shape closely resembles the lensing $B$-mode spectrum. Introducing $a$ allows us to separate the true tensor contribution from the wedge-induced contamination and to assess to what extent this additional degree of freedom mitigates the bias in the recovered value of $r$.}
\simone{Figure~\ref{fig:quad_fit_2d} presents the results of the two-parameter fit introduced above.  The model including the additional parameter $a$ provides a noticeably improved description of the spectra compared to the single-parameter case shown in Figure~\ref{fig:quad_fit}, confirming that the wedge-induced contamination cannot be fully absorbed by rescaling the primordial tensor contribution alone. As the wedge angle increases, the fitted value $a_{\rm fit}$ systematically grows, indicating that the wedge introduces excess power with a lensing-like spectral shape. This behaviour is consistent with previous studies showing that pointing-related and polarization-angle systematics can produce spurious $B$-mode power that mimics lensing and bias the lensing reconstruction~\cite{litebird2023probing,2021Mirmelstein,2009Su}.}
\subsection{Scaling of the wedge-like effect with the number of detectors}
\label{sec:four}
\begin{figure}[t]
\centering
\includegraphics[width=0.75\textwidth]{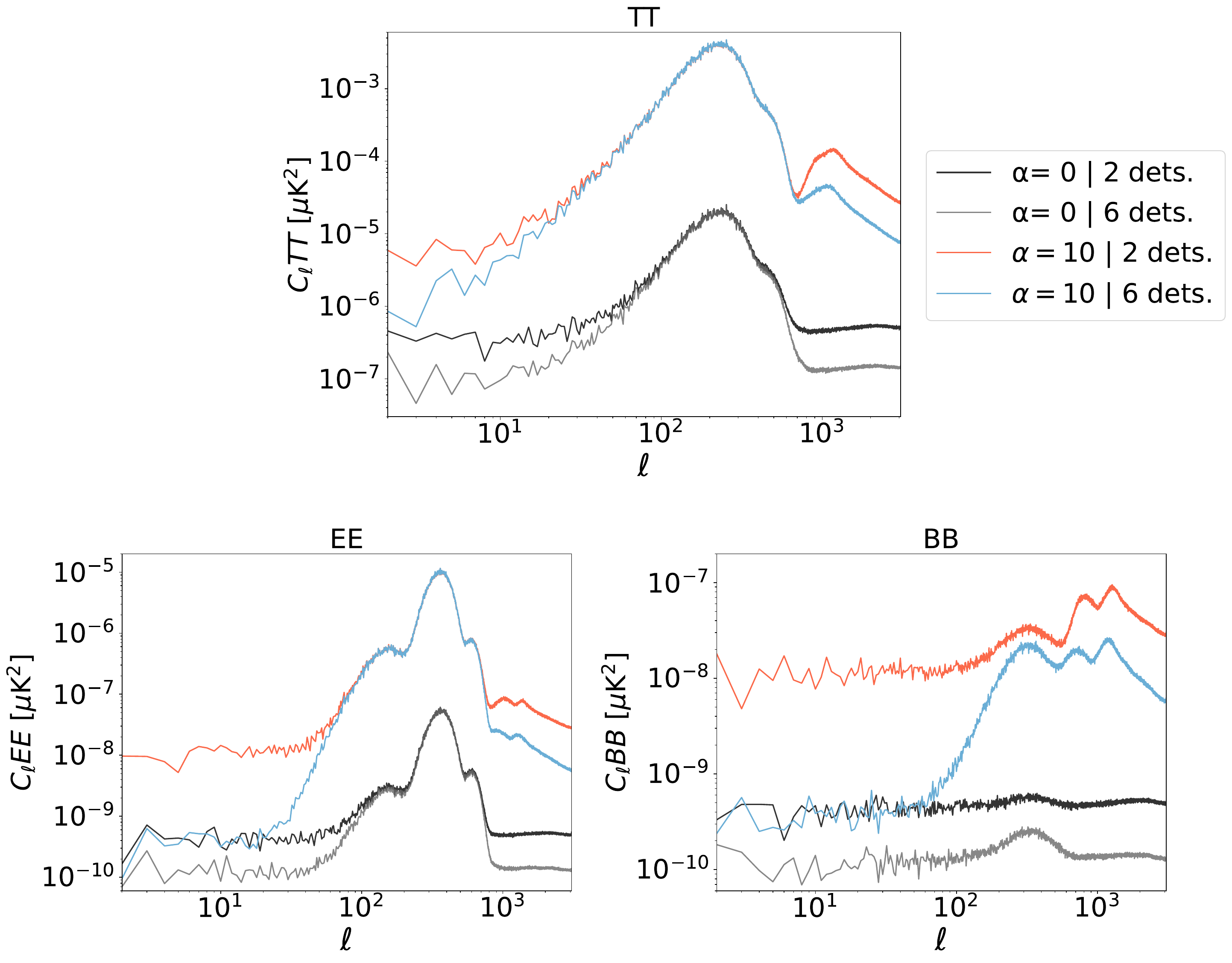}
\caption{Angular power spectra of residual maps for $TT$, $EE$, and $BB$ components, showing the impact of different wedge angles (\(\alpha\)) and detector configurations. The cases are represented as follows: input spectrum (light gray), \(\alpha = 0\)\,arcmin with two detectors (black) and six detectors (dark gray), and \(\alpha = 10\)\,arcmin with two detectors (red) and six detectors (blue).}
    \label{fig:quad_diff_ang_spectra_output}
\end{figure}

An additional, particularly insightful step in our analysis is to assess how the wedge-like systematics changes when increasing the number of detectors. We maintain all parameters used this far, with the sole adjustment being an increase in the number of detectors from two to four and then to six. We selected orthogonal pairs of detectors, widely angularly separated on the focal plane, ensuring that each pair individually covered the entire sky. \simone{This setup is deliberately simplified and far from the final \LiteBIRD configuration, which will employ between 36 and approximately 750 detectors per frequency band~\cite{litebird2023probing}; here the purpose is to illustrate the scaling behaviour rather than reproduce the full instrument}. The results are consistent with those previously obtained, but they generally show less intense residuals in both maps and spectra. As an example, Figure~\ref{fig:quad_diff_ang_spectra_output} shows the residual spectra for configurations with \(\alpha = [0, 10]\) arcminutes, comparing cases with two and six detectors. Note that where not explicitly shown, it is implicitly understood that the units are in arcminutes.
By comparing cases with two, four and six detectors, we observe in Table~\ref{tab:summary} that, as the number of detector pairs increases, the requirements become less stringent. Similarly, Table~\ref{tab:quad_vs_a2} presents a comparison of analysis results analogous to those in Section~\ref{sec:fake}, for two, four, and six detectors. Compared to the scenario shown in Section~\ref{sec:fake}, we observe generally lower \(a_\text{{fit}}\) values as the number of detector pairs increases.\\
\simone{It is worth noting that for large misalignments, the dominant errors shift from wedge-induced pointing disturbances to systematic effects associated with beam reconstruction. In this regime, increasing the number of detectors in simulations does not improve the robustness of the wedge constraint. We therefore restrict our analysis to the small-angle regime, where the wedge is the leading systematic and its scaling with the number of detectors can be meaningfully assessed. To gain an intuition of the threshold misalignment, we compare our case to the transmissive HWP configuration discussed in~\cite{litebird2023probing}.}
\simone{As reported in Section 5.3.7.3 of~\cite{litebird2023probing}, for a \emph{transmissive} sapphire rotating HWP the maximal allowed wedge angle $\psi$ was constrained to be $\psi<4$ arcmin to satisfy the benchmark $\Delta r_{\text{max}} = 5.7\times 10^{-6}$. In that geometry, a wedge produces a boresight pointing deviation}
\begin{equation}
\label{eq:phi_trans}
\phi_{\rm trans} \simeq (n-1)\,\psi,
\end{equation}
\simone{where $\phi$ is the pointing deviation and $n$ is the refractive index of the plate material (e.g.\ $n\simeq 3.1$ for sapphire). In the \emph{reflective} case considered here, the wedge-like tilt of the modulator produces a purely geometrical pointing deviation (see Figure~\ref{fig:wedge})}
\begin{equation}
\phi_{\rm refl} = 2\,\alpha,
\end{equation}
\simone{where $\alpha$ is the modulator tilt (wedge angle) defined in Section~\ref{sec:wedge_intro}.
Equating the two deviations to compare tolerances gives}
\begin{equation}
\alpha_{\rm eq} \simeq \frac{n-1}{2}\,\psi.
\end{equation}
\simone{Using the transmissive benchmark $\psi = 4$ arcmin and $n \simeq 3.1$ gives $\alpha_{\rm eq} \approx 4.2$ arcmin. Here, $\alpha_{\rm eq}$ denotes the tilt angle that reproduces the same pointing deviation as in the transmissive case.}
\simone{The \LiteBIRD HWP is being considered as a transmissive meta-material component in the ongoing redesign process.
The expected meta-material phase-shift produces the same effect of a material with refractive index $n_\text{meta}$ such that $ 1.9 \le n_\text{meta} \le 2.3$ within the operating band. The equivalent maximum wedge angle ($\psi_{\text{meta}}$) can be derived imposing the same pointing deviation, resulting in a requirement -- in the most stringent case, $n_\text{meta} = 2.3$,}
\begin{equation}
    \psi_{\text{meta}} = \frac{\phi_{\rm trans}}{n_{\text{meta}}-1} = 6.5 \text{ arcmin}
\end{equation}
\simone{This value demonstrates a modest but significant relaxation of the requirement when a meta-material with an equivalent lower index of refraction is employed.}

\section{Conclusions}
\label{sec:conclusions}
This paper has focused on studying the systematic effects caused by the misalignment of the HWP due to a constant wedge angle in the MHFT onboard the \LiteBIRD mission.

\paragraph{Maximum wedge angle and error analysis.} One of the key results of this study is the maximum acceptable wedge angle of the HWP, $\alpha_{\max }$, and the corresponding maximum error in the mechanical mounting, $\sigma_x$, beyond which the systematic errors introduced by the wedge-like effect become unacceptably large for the scientific objectives of the \LiteBIRD mission.\\
We summarize the results for an HWP with diameter of 500\,mm in Table~\ref{tab:summary}.\\
This level of precision underscores the critical need for highly accurate alignment mechanisms in the \LiteBIRD instrument and as we can see from the table by increasing the number of detectors we get less stringent requirements.
\begin{table}[t]
\centering
\begin{tabular}{c| c c}
\hline
N dets. & \textbf{$\alpha_\text{{max}}$}\,[\text{arcmin}] & $\sigma_x$\,[\text{mm}] \\
\hline
2 & 12.7 & 2 \\
4 & 22.4 & 3 \\
6 & 22.7 & 3 \\
\hline
\end{tabular}
\caption{Summary of results for the maximum allowable wedge angle $\alpha_{\max }$ and respective vertical edge displacement $\sigma_x$, obtained with \( C_{\ell} = r C_{\ell}^{\text{tens}} + C_{\ell}^{\text{lens}} \) fit model, for an HWP with diameter of $d=500$\,mm.}
\label{tab:summary}
\end{table}
However, it is important to clarify that these requirements are derived by interpreting the entire systematic-induced signal as a spurious contribution to $r$. As shown in Section~\ref{sec:dr}, this is not a reasonable approach: the shape of the angular power spectra induced by the wedge differs significantly from that of a primordial tensor signal, leading to an artificial increase of $r_{\text{fit}}$ as a function of the wedge angle $\alpha$.

\paragraph{Impact of the wedge-like effect on $B$-mode detection.} The simulations also demonstrated the effect of the wedge angle on the detection of the primordial $B$ modes. Even small misalignments in the HWP can introduce spurious signals into the TOD that propagate through the entire data processing chain, leading to distortions in the reconstructed polarization maps and the angular power spectra. One particular effect is the generation of a spurious signal that mimics the effects of gravitational lensing. This contamination could hinder the mission's ability to accurately constrain the tensor-to-scalar ratio $r$, which is directly related to the strength of primordial gravitational waves. By increasing the number of detectors, we observe lower level of the spurious lensing signal and as a consequence lower $a_\text{{fit}}$ values for $\alpha > 0$, where the parameter $a_\text{{fit}}$ quantifies the additional fake lensing caused by the wedge. We summarize the analysis results in Table~\ref{tab:quad_vs_a2}. Typically for CMB experiments, the noise of a single detector is high compared to the $B$-mode lensing signal. Therefore, the wedge does not create significant issues, other than a slight increase of the lensing signal. This can also be interpreted as an increase in the detector white noise. It would be helpful in future studies to run simulation with a very large number of detectors and noise to estimate if the cancellation due to the combination of many detectors scales linearly with the number of detectors or if it eventually reaches a saturation point.\\
In conclusion, this work has provided new insights into the impact of the wedge-like effect on the \LiteBIRD mission. By establishing precise alignment requirements and quantifying the systematic errors associated with different wedge angles, this paper contributes to ensuring that the mission's scientific goals can be achieved. With careful attention to these systematic effects, \LiteBIRD will be well-positioned to make groundbreaking discoveries about the early universe and the physics of cosmic inflation.

\begin{table}[t]
\centering
\begin{tabular}{c| c| c c c}
\hline
&N dets.  & \textbf{$\alpha=0$} & \textbf{$\alpha=5$} & \textbf{$\alpha=10$} \\
\hline
\multirow{3}{*}{$a_\text{{fit}}$} & 2 & 0.000 & 0.003 & 0.010 \\
                                  & 4 & 0.000 & 0.000 & 0.004 \\
                                  & 6 & 0.000 & 0.000 & 0.004 \\

\hline
\end{tabular}
\caption{Summary of results for $a_\text{{fit}}$, obtained with \(C_\ell = r\,C_\ell^{\rm tens} + (A_L + a)\,C_\ell^{\rm lens}\) fit model. \simone{In this fitting procedure, $r_{\rm fit}$ consistently returns zero within numerical precision, so only the fitted values of $a$ are reported, as the wedge–induced signal in this setup projects primarily as a lensing–like contribution.}}
\label{tab:quad_vs_a2}
\end{table}

\paragraph{Exploring wedge angle variability in future studies.} Our analysis assumed a constant wedge angle throughout this work. However, the potential wobbling of the reflective HWP introduces a time-dependent variation of the wedge angle, which could affect the derived constraints. In fact the constraint derived for the constant case may not directly apply if the wedge angle varies stochastically or periodically over time. Specifically, a time-dependent $\alpha$ could lead to either an overestimation or underestimation of the effective wedge angle compared to the static case. Furthermore, if different values of $\alpha$ are used in the TOD production and in the map-making process, systematic errors could be introduced, potentially biasing the final results. In addition, future work will also consider more realistic scenarios where detectors within each pair are not perfectly balanced, as such imbalance could contribute to additional systematic effects.
A detailed investigation of these effects is left for future work.

\acknowledgments
S.Mi. is supported by European Union - Next Generation EU,  Missione: I.4.1 Borse dottorati generici ricerca PNRR (Missione 4),  Componente: 1, CUP 351: B83C22003210006. 
The Italian \LiteBIRD phase A contribution is supported by the Italian Space Agency (ASI Grants No. 2020-9-HH.0 and 2016-24-H.1-2018), the National Institute for Nuclear Physics (INFN) and the National Institute for Astrophysics (INAF). This work has also received funding by the European Union’s Horizon 2020 research and innovation program under grant agreement no. 101007633 CMB-Inflate.


\bibliographystyle{JHEP}
\bibliography{references.bib}

\end{document}